\documentstyle[10pt,epsf,epsfig,rotating,amssymb,dp_delphititle]{dp_delphi}
\makeindex
\def\DpPaperGroup{EP}
\def\DpPaperRef{2000-015}
\def\DpDate{20 January 2000}
\def\DpAuthors{DELPHI Collaboration}
\def\DpSubmit{(Eur. Phys. J. C16(2000)211)}
\def\DpTitle{{ Search for supersymmetric particles in
 scenarios with a gravitino LSP and stau NLSP}}
\def\DpComment{ }
\def\DpEMail{ }
\newcommand{\stau} {$\tilde{\tau}$}
\newcommand{\stuno} {$\tilde{\tau}_1$}
\newcommand{\staur} {$\tilde{\tau}_R$}
\newcommand{\staul} {$\tilde{\tau}_L$}
\newcommand{\selr} {$\tilde{e}_R$}
\newcommand{\smur} {$\tilde{\mu}_R$}
\newcommand{\slep} {$\tilde{l}$}
\newcommand{\nuno} {$\tilde{\chi}^0_1$}
\newcommand{\chino} {$\tilde{\chi}^\pm_1$}
\newcommand{\grav} {$\tilde{G}$}
\newcommand{\mgrav} {$m_{\tilde{G}}$}
\newcommand{\ra} {\rightarrow}
\newcommand{\eeto} {\mbox{$ {\mathrm e}^+ {\mathrm e}^-\! \ra\ $}}
\newcommand{\ee} {\mbox{$ {\mathrm e}^+ {\mathrm e}^-$}}
\newcommand{\qqbar} {$q\bar{q}$}
\newcommand{\Wp} {\mbox{$ {\mathrm W}^+$}}
\newcommand{\Wm} {\mbox{$ {\mathrm W}^-$}}
\newcommand{\Zn} {\mbox{$ {\mathrm Z}$}}
\newcommand{\Wev} {\mbox{$ {\mathrm{W e}} \nu_{\mathrm e}$}}

\newcommand{\Zee} {\mbox{$ \Zn \ee$}}
\newcommand{\GeV} {~\mbox{${\mathrm{GeV}}$}}
\newcommand{\GeVcc} {~\mbox{${\mathrm{GeV}}/c^2$}}
\newcommand{\eVcc} {~\mbox{${\mathrm{eV}}/c^2$}}

\newcommand{\GeVc} {~\mbox{$ {\mathrm{GeV}}/c $}}

\newcommand{\TeV} {~\mbox{$ {\mathrm{TeV}} $}}
\newcommand{\keVcc} {~\mbox{$ {\mathrm{keV}}/{\mathrm{c}}^2 $}}
\newcommand{\etal} {\mbox{\it et al.}}
\def\NPB#1#2#3{{\rm Nucl.~Phys.} {\bf{B#1}} (19#2) #3}
\def\PLB#1#2#3{{\rm Phys.~Lett.} {\bf{B#1}} (19#2) #3}
\def\PRD#1#2#3{{\rm Phys.~Rev.} {\bf{D#1}} (19#2) #3}
\def\PRL#1#2#3{{\rm Phys.~Rev.~Lett.} {\bf{#1}} (19#2) #3}
\def\ZPC#1#2#3{{\rm Z.~Phys.} {\bf C#1} (19#2) #3}

\def\NIMA#1#2#3{{\rm Nucl.~Instr.~and~Meth.} {\bf#1} (19#2) #3}
\def\CPC#1#2#3{{\rm Comp.~Phys.~Comm.} {\bf#1} (19#2) #3}
\begin{document}
\makeatletter
\newcount\@tempcntc
\def\@citex[#1]#2{\if@filesw\immediate\write\@auxout{\string\citation{#2}}\fi
\@tempcnta\z@\@tempcntb\m@ne\def\@citea{}\@cite{\@for\@citeb:=#2\do
{\@ifundefined
{b@\@citeb}{\@citeo\@tempcntb\m@ne\@citea\def\@citea{,}{\bf ?}\@warning
{Citation `\@citeb' on page \thepage \space undefined}}%
{\setbox\z@\hbox{\global\@tempcntc0\csname b@\@citeb\endcsname\relax}%
\ifnum\@tempcntc=\z@ \@citeo\@tempcntb\m@ne
\@citea\def\@citea{,}\hbox{\csname b@\@citeb\endcsname}%
\else
\advance\@tempcntb\@ne
\ifnum\@tempcntb=\@tempcntc
\else\advance\@tempcntb\m@ne\@citeo
\@tempcnta\@tempcntc\@tempcntb\@tempcntc\fi\fi}}\@citeo}{#1}}
\def\@citeo{\ifnum\@tempcnta>\@tempcntb\else\@citea\def\@citea{,}%
\ifnum\@tempcnta=\@tempcntb\the\@tempcnta\else
{\advance\@tempcnta\@ne\ifnum\@tempcnta=\@tempcntb \else \def\@citea{--}\fi
\advance\@tempcnta\m@ne\the\@tempcnta\@citea\the\@tempcntb}\fi\fi}
\makeatother
\begin{titlepage}
\pagenumbering{roman}
\CERNpreprint{\DpPaperGroup}{\DpPaperRef} 
\date{{\small\DpDate}} 
\title{\DpTitle} 
\address{\DpAuthors} 
\begin{shortabs} 
\noindent
%
\noindent

Sleptons, neutralinos and charginos
were searched for in the context of scenarios where the lightest 
supersymmetric particle is the gravitino.
It was assumed that the stau is the next-to-lightest 
supersymmetric particle.
Data collected with the DELPHI
detector at a centre-of-mass energy near 189 GeV 
were analysed combining 
the methods developed in previous searches at lower energies.
No evidence 
for the production of these supersymmetric particles was found.
Hence, 
limits were derived at 95\% confidence level.
%
%
\noindent
\end{shortabs}
\vfill
\begin{center}
\DpSubmit \ \\ 
\DpComment \ \\
\DpEMail \ \\
\end{center}
\vfill
\clearpage
\headsep 10.0pt
\addtolength{\textheight}{10mm}
\addtolength{\footskip}{-5mm}
\begingroup
%
\newcommand{\DpName}[2]{\hbox{#1$^{\ref{#2}}$},\hfill}
\newcommand{\DpNameTwo}[3]{\hbox{#1$^{\ref{#2},\ref{#3}}$},\hfill}
\newcommand{\DpNameThree}[4]{\hbox{#1$^{\ref{#2},\ref{#3},\ref{#4}}$},\hfill}
\newskip\Bigfill \Bigfill = 0pt plus 1000fill
\newcommand{\DpNameLast}[2]{\hbox{#1$^{\ref{#2}}$}\hspace{\Bigfill}}
%
\footnotesize
\noindent
\DpName{P.Abreu}{LIP}
\DpName{W.Adam}{VIENNA}
\DpName{T.Adye}{RAL}
\DpName{P.Adzic}{DEMOKRITOS}
\DpName{Z.Albrecht}{KARLSRUHE}
\DpName{T.Alderweireld}{AIM}
\DpName{G.D.Alekseev}{JINR}
\DpName{R.Alemany}{VALENCIA}
\DpName{T.Allmendinger}{KARLSRUHE}
\DpName{P.P.Allport}{LIVERPOOL}
% ...
\normalsize
\endgroup
\titlefoot{Department of Physics and Astronomy, Iowa State
     University, Ames IA 50011-3160, USA
    \label{AMES}}
\titlefoot{Physics Department, Univ. Instelling Antwerpen,
     Universiteitsplein 1, B-2610 Antwerpen, Belgium \\
     \indent~~and IIHE, ULB-VUB,
     Pleinlaan 2, B-1050 Brussels, Belgium \\
     \indent~~and Facult\'e des Sciences,
     Univ. de l'Etat Mons, Av. Maistriau 19, B-7000 Mons, Belgium
    \label{AIM}}
\titlefoot{Physics Laboratory, University of Athens, Solonos Str.
     104, GR-10680 Athens, Greece
    \label{ATHENS}}
\titlefoot{Department of Physics, University of Bergen,
     All\'egaten 55, NO-5007 Bergen, Norway
    \label{BERGEN}}
\titlefoot{Dipartimento di Fisica, Universit\`a di Bologna and INFN,
     Via Irnerio 46, IT-40126 Bologna, Italy
    \label{BOLOGNA}}
\titlefoot{Centro Brasileiro de Pesquisas F\'{\i}sicas, rua Xavier Sigaud 150,
     BR-22290 Rio de Janeiro, Brazil \\
     \indent~~and Depto. de F\'{\i}sica, Pont. Univ. Cat\'olica,
     C.P. 38071 BR-22453 Rio de Janeiro, Brazil \\
     \indent~~and Inst. de F\'{\i}sica, Univ. Estadual do Rio de Janeiro,
     rua S\~{a}o Francisco Xavier 524, Rio de Janeiro, Brazil
    \label{BRASIL}}
\titlefoot{Comenius University, Faculty of Mathematics and Physics,
     Mlynska Dolina, SK-84215 Bratislava, Slovakia
    \label{BRATISLAVA}}
\titlefoot{Coll\`ege de France, Lab. de Physique Corpusculaire, IN2P3-CNRS,
     FR-75231 Paris Cedex 05, France
    \label{CDF}}
\titlefoot{CERN, CH-1211 Geneva 23, Switzerland
    \label{CERN}}
\titlefoot{Institut de Recherches Subatomiques, IN2P3 - CNRS/ULP - BP20,
     FR-67037 Strasbourg Cedex, France
    \label{CRN}}
\titlefoot{Now at DESY-Zeuthen, Platanenallee 6, D-15735 Zeuthen, Germany
    \label{DESY}}
\titlefoot{Institute of Nuclear Physics, N.C.S.R. Demokritos,
     P.O. Box 60228, GR-15310 Athens, Greece
    \label{DEMOKRITOS}}
\titlefoot{FZU, Inst. of Phys. of the C.A.S. High Energy Physics Division,
     Na Slovance 2, CZ-180 40, Praha 8, Czech Republic
    \label{FZU}}
\titlefoot{Dipartimento di Fisica, Universit\`a di Genova and INFN,
     Via Dodecaneso 33, IT-16146 Genova, Italy
    \label{GENOVA}}
\titlefoot{Institut des Sciences Nucl\'eaires, IN2P3-CNRS, Universit\'e
     de Grenoble 1, FR-38026 Grenoble Cedex, France
    \label{GRENOBLE}}
\titlefoot{Helsinki Institute of Physics, HIP,
     P.O. Box 9, FI-00014 Helsinki, Finland
    \label{HELSINKI}}
\titlefoot{Joint Institute for Nuclear Research, Dubna, Head Post
     Office, P.O. Box 79, RU-101 000 Moscow, Russian Federation
    \label{JINR}}
\titlefoot{Institut f\"ur Experimentelle Kernphysik,
     Universit\"at Karlsruhe, Postfach 6980, DE-76128 Karlsruhe,
     Germany
    \label{KARLSRUHE}}
\titlefoot{Institute of Nuclear Physics and University of Mining and Metalurgy,
     Ul. Kawiory 26a, PL-30055 Krakow, Poland
    \label{KRAKOW}}
\titlefoot{Universit\'e de Paris-Sud, Lab. de l'Acc\'el\'erateur
     Lin\'eaire, IN2P3-CNRS, B\^{a}t. 200, FR-91405 Orsay Cedex, France
    \label{LAL}}
\titlefoot{School of Physics and Chemistry, University of Lancaster,
     Lancaster LA1 4YB, UK
    \label{LANCASTER}}
\titlefoot{LIP, IST, FCUL - Av. Elias Garcia, 14-$1^{o}$,
     PT-1000 Lisboa Codex, Portugal
    \label{LIP}}
\titlefoot{Department of Physics, University of Liverpool, P.O.
     Box 147, Liverpool L69 3BX, UK
    \label{LIVERPOOL}}
\titlefoot{LPNHE, IN2P3-CNRS, Univ.~Paris VI et VII, Tour 33 (RdC),
     4 place Jussieu, FR-75252 Paris Cedex 05, France
    \label{LPNHE}}
\titlefoot{Department of Physics, University of Lund,
     S\"olvegatan 14, SE-223 63 Lund, Sweden
    \label{LUND}}
\titlefoot{Universit\'e Claude Bernard de Lyon, IPNL, IN2P3-CNRS,
     FR-69622 Villeurbanne Cedex, France
    \label{LYON}}
\titlefoot{Univ. d'Aix - Marseille II - CPP, IN2P3-CNRS,
     FR-13288 Marseille Cedex 09, France
    \label{MARSEILLE}}
\titlefoot{Dipartimento di Fisica, Universit\`a di Milano and INFN-MILANO,
     Via Celoria 16, IT-20133 Milan, Italy
    \label{MILANO}}
\titlefoot{Dipartimento di Fisica, Univ. di Milano-Bicocca and
     INFN-MILANO, Piazza delle Scienze 2, IT-20126 Milan, Italy
    \label{MILANO2}}
\titlefoot{Niels Bohr Institute, Blegdamsvej 17,
     DK-2100 Copenhagen {\O}, Denmark
    \label{NBI}}
\titlefoot{IPNP of MFF, Charles Univ., Areal MFF,
     V Holesovickach 2, CZ-180 00, Praha 8, Czech Republic
    \label{NC}}
\titlefoot{NIKHEF, Postbus 41882, NL-1009 DB
     Amsterdam, The Netherlands
    \label{NIKHEF}}
\titlefoot{National Technical University, Physics Department,
     Zografou Campus, GR-15773 Athens, Greece
    \label{NTU-ATHENS}}
\titlefoot{Physics Department, University of Oslo, Blindern,
     NO-1000 Oslo 3, Norway
    \label{OSLO}}
\titlefoot{Dpto. Fisica, Univ. Oviedo, Avda. Calvo Sotelo
     s/n, ES-33007 Oviedo, Spain
    \label{OVIEDO}}
\titlefoot{Department of Physics, University of Oxford,
     Keble Road, Oxford OX1 3RH, UK
    \label{OXFORD}}
\titlefoot{Dipartimento di Fisica, Universit\`a di Padova and
     INFN, Via Marzolo 8, IT-35131 Padua, Italy
    \label{PADOVA}}
\titlefoot{Rutherford Appleton Laboratory, Chilton, Didcot
     OX11 OQX, UK
    \label{RAL}}
\titlefoot{Dipartimento di Fisica, Universit\`a di Roma II and
     INFN, Tor Vergata, IT-00173 Rome, Italy
    \label{ROMA2}}
\titlefoot{Dipartimento di Fisica, Universit\`a di Roma III and
     INFN, Via della Vasca Navale 84, IT-00146 Rome, Italy
    \label{ROMA3}}
\titlefoot{DAPNIA/Service de Physique des Particules,
     CEA-Saclay, FR-91191 Gif-sur-Yvette Cedex, France
    \label{SACLAY}}
\titlefoot{Instituto de Fisica de Cantabria (CSIC-UC), Avda.
     los Castros s/n, ES-39006 Santander, Spain
    \label{SANTANDER}}
\titlefoot{Dipartimento di Fisica, Universit\`a degli Studi di Roma
     La Sapienza, Piazzale Aldo Moro 2, IT-00185 Rome, Italy
    \label{SAPIENZA}}
\titlefoot{Inst. for High Energy Physics, Serpukov
     P.O. Box 35, Protvino, (Moscow Region), Russian Federation
    \label{SERPUKHOV}}
\titlefoot{J. Stefan Institute, Jamova 39, SI-1000 Ljubljana, Slovenia
     and Laboratory for Astroparticle Physics,\\
     \indent~~Nova Gorica Polytechnic, Kostanjeviska 16a, SI-5000 Nova Gorica, Slovenia, \\
     \indent~~and Department of Physics, University of Ljubljana,
     SI-1000 Ljubljana, Slovenia
    \label{SLOVENIJA}}
\titlefoot{Fysikum, Stockholm University,
     Box 6730, SE-113 85 Stockholm, Sweden
    \label{STOCKHOLM}}
\titlefoot{Dipartimento di Fisica Sperimentale, Universit\`a di
     Torino and INFN, Via P. Giuria 1, IT-10125 Turin, Italy
    \label{TORINO}}
\titlefoot{Dipartimento di Fisica, Universit\`a di Trieste and
     INFN, Via A. Valerio 2, IT-34127 Trieste, Italy \\
     \indent~~and Istituto di Fisica, Universit\`a di Udine,
     IT-33100 Udine, Italy
    \label{TU}}
\titlefoot{Univ. Federal do Rio de Janeiro, C.P. 68528
     Cidade Univ., Ilha do Fund\~ao
     BR-21945-970 Rio de Janeiro, Brazil
    \label{UFRJ}}
\titlefoot{Department of Radiation Sciences, University of
     Uppsala, P.O. Box 535, SE-751 21 Uppsala, Sweden
    \label{UPPSALA}}
\titlefoot{IFIC, Valencia-CSIC, and D.F.A.M.N., U. de Valencia,
     Avda. Dr. Moliner 50, ES-46100 Burjassot (Valencia), Spain
    \label{VALENCIA}}
\titlefoot{Institut f\"ur Hochenergiephysik, \"Osterr. Akad.
     d. Wissensch., Nikolsdorfergasse 18, AT-1050 Vienna, Austria
    \label{VIENNA}}
\titlefoot{Inst. Nuclear Studies and University of Warsaw, Ul.
     Hoza 69, PL-00681 Warsaw, Poland
    \label{WARSZAWA}}
\titlefoot{Fachbereich Physik, University of Wuppertal, Postfach
     100 127, DE-42097 Wuppertal, Germany
    \label{WUPPERTAL}}
\addtolength{\textheight}{-10mm}
\addtolength{\footskip}{5mm}
\clearpage
\headsep 30.0pt
\end{titlepage}
%
\pagenumbering{arabic} 
\setcounter{footnote}{0} %
\large
%
\section{Introduction}
\label{sec:intro}
Supersymmetry (SUSY) may be broken at a scale below the 
grand-unification scale $M_{GUT}$, with the ordinary gauge interactions
acting as the messengers of supersymmetry breaking \cite{Dine1,Dine2}.
In the corresponding models (GMSB models), 
the gravitino, $\tilde{G}$, turns out
to be the lightest supersymmetric particle
(LSP) and is expected to be almost massless. The next-to-lightest 
supersymmetric particle (NLSP) is therefore unstable and decays, 
under the assumption of $R$-parity conservation, into 
its ordinary matter partner and an invisible 
gravitino.

The number of generations of supersymmetry breaking messengers 
and the value of $\tan\beta$\ 
usually determine which supersymmetric particle is the 
NLSP~\cite{Bagger,Dutta,francesca,giudice}. In the majority of the GMSB 
space, the NLSP is a slepton, \slep. Moreover, depending on 
magnitude of the mixing
between \staur~and \staul, there exist two possible 
scenarios. If the mixing is large~\footnote{In GMSB models large mixing 
occurs generally in regions of $\tan\beta\geq 1$0 or $|\mu|>50$0 GeV.}, 
\stuno\ is the NLSP, but if the mixing
is negligible, \stuno\ is mainly right-handed~\cite{bartl}
and almost mass degenerate with the other sleptons. In this case, the
\selr\ and \smur\ three 
body decay (\slep $\rightarrow$ \stuno $\tau l$ with
\stuno $\rightarrow~\tau$ \grav), is very suppressed, and \selr~and
\smur~decay directly into $l$\grav. This scenario is called \slep~co-NLSP.  
Searches for supersymmetric particles within 
both these scenarios are described in this article.

Due to the coupling of the NLSP to \grav, its mean decay length 
can range from $\mu m$ to meters depending on the mass of the gravitino 
(\mgrav), or equivalently, on the
scale of SUSY breaking, $\sqrt{F}$:
\begin{equation}
L = 1.76 \times 10^{-3} \sqrt{\left (\frac{E_{\tilde{l}}}{m_{\tilde{l}}}
\right )^2-1}
\left ( \frac{m_{\tilde{l}}}{100 \, {\rm GeV/c}^2} \right )^{-5}
\left ( \frac{m_{\tilde{G}}}{1\, \rm eV/c^2}\right )^{2} \;\; {\mathrm cm,}
\label{life}
\end{equation}

For example, for $m_{\tilde{G}}\lesssim  250$ eV
($\sqrt{F} \lesssim$ 1000 \TeV), the decay of the NLSP can take place
within the detector. This range of $\sqrt{F}$ is in fact consistent with
astrophysical and cosmological considerations \cite{Dinopoulos0,wagner}.


The results of three searches are presented in this work.
The first one looks for the production of \nuno~pairs with either
$\tilde{\chi}^0_1$ decaying to $\tilde{\tau}_1 \tau$\ and 
$\tilde{\tau}_1$ then decaying promptly into
$\tau \tilde{G}$, which is an update of the search presented in 
ref.~\cite{nuestro_papel}, or 
$\tilde{\chi}^0_1$ decaying to $\tilde{l} l$\ 
with BR($\tilde{\chi}^0_1 \to \tilde{l} l$)~=~1/3 and 
$\tilde{l}$\  promptly decaying into
$l \tilde{G}$: $e^+e^- \to \tilde{\chi}^0_1 \tilde{\chi}^0_1 \to 
\tilde{l} l \tilde{l'} l'  \to l \tilde{G} l l' \tilde{G} l'$.
These two modes represent the two extremes in the range of possible decays of
the neutralino.
In particular, a higgsino-like \nuno~ 
would decay only to $\tilde{\tau}_1 \tau$\ for all practical purposes since 
the higgsino component of the \nuno~ couples to \slep~through Yukawa couplings.
On the other hand, the decays of a gaugino-like \nuno~are regulated only by
phase space considerations. Therefore, in the case of the \stuno-NLSP scenario,
neutralino pair production would mainly lead to 
a final state with four tau leptons and two gravitinos, while 
in the case of a co-NLSP
scenario, the final signature would contain two pairs of 
leptons with possibly different flavour and two gravitinos.


The second search concerns $\tilde{l}$ pair production followed
by the decays $\tilde{l} \rightarrow l \tilde{G}$
within the detector volume.
The signature of such an event will be at least one
track of a charged particle
with a kink or a decay vertex when the
$\tilde{l}$ decays inside the tracking devices. If the decay length is
too short (small $m_{\tilde{G}}$) to allow for the reconstruction of the
$\tilde{l}$ track, only the corresponding lepton or its decay products 
will be seen in
the detector, and the search will then be based on track impact parameter.
However, if the decay takes place outside the tracking devices
(large $m_{\tilde{G}}$),
the signature will be that of a heavy charged particle already studied
by DELPHI~\cite{heavyparticles}. 
For very light gravitinos the limits
from the search for sleptons in gravity mediated (MSUGRA) models 
can be applied~\cite{slep_189}.


The third search looks for the pair-production of lightest charginos, \chino.
Charginos, if produced in this context, would promptly decay through the
channel $\tilde{\chi}_1^+ \ra \tilde{\tau}_1^+\nu_{\tau}$\ \cite{chargmsb}.
The \stuno~would then decay 
into $\tau$\grav, with non negligible mean lifetime.
This search is divided into three sub-channels according to the mean lifetime 
of the stau as explained in the previous paragraph: two acoplanar leptons 
with missing energy, at least one track with large impact parameter or a 
kink, or at least one track corresponding to a very massive stable particle.


The data samples and event selections are respectively described in 
sections~\ref{experimentalprocedure} 
and~\ref{dataselection}, while the results and a model dependent
interpretation
are presented in
section~\ref{sec:resultados}. 
%
%
\section{Event sample and experimental procedure}
\label{experimentalprocedure}

All searches are based on 
data collected with the DELPHI detector
 during 1998 at a
centre-of-mass energy of 189 GeV.
The total integrated luminosity was 153.6~${\mathrm pb}^{-1}$.  
A detailed
description of the DELPHI detector can be found in \cite{detector} and its
performance in \cite{performance}.
 In all cases, the \stuno -NLSP scenario
searches are updates to similar searches carried out at lower 
centre-of-mass energies. All co-NLSP scenario searches are carried out 
at $\sqrt{s}$ = 189 GeV.

To evaluate the signal efficiencies and Standard Model (SM) background 
contaminations,
events were ge\-ne\-ra\-ted using different programs, all
relying on {\tt JETSET} 7.4 \cite{JETSET}, tuned
to LEP~1 data \cite{TUNE} for quark fragmentation.
The program {\tt SUSYGEN} \cite{SUSYGEN} was used to generate
the neutralino pair events and their subsequent decay products. 
In order to compute detection efficiencies, a
total of 42000 events were generated 
with masses 
67\GeVcc$\leq m_{\tilde{\tau}_1}+2$\GeVcc~$\leq m_{\tilde{\chi}^0_1} \leq 
  \sqrt{s}/2$.
A \stau\ pair sample of 36000 events 
(subdivided in 36 samples) was produced 
with {\tt PYTHIA} 5.7 \cite{JETSET} with
staus having mean decay lengths from  0.25 to 200 cm and masses
from $m_{\tau}$ to 90\GeVcc.
Another sample of \stau\ pair was produced with {\tt SUSYGEN} for the 
small impact parameter search with $m_{\tilde{\tau}_1}$\ from 7 to 80\GeVcc.
Similar samples of smuons and selectrons were produced to study the 
sleptons co-NLSP scenario.

{\tt SUSYGEN} was also used to generate
the \chino\ pair production samples and their decays. 
In order to compute detection efficiencies, a
total of 45 samples of 500 events each were generated 
with \grav~masses of 1, 100 and 1000\eVcc ,
$ m_{\tilde{\tau}_1}+ 0.3$\GeVcc~$\leq m_{\tilde{\chi}^+_1} \leq 
\sqrt{s}/2$\ and
 $m_{\tilde{\tau}_1}\geq 65$\GeVcc. Samples with smaller $\Delta m = 
m_{\tilde{\chi}^+_1} -m_{\tilde{\tau}_1}$\ were not generated because 
in this region the \chino\ does not decay mainly to 
$\tilde{\tau}_1$\ and $\nu_{\tau}$ but into W and $\tilde{G}$.
The different background samples and event selections are described in 
references~\cite{chargino_183,chargino_189,heavyparticles} 
for $m_{\tilde{G}} = $1 and 
1000\eVcc\ respectively. For the case of $m_{\tilde{G}} = 100$\eVcc , the
analysis is the same as the one used in the search for sleptons of this paper
and consequently, the same sample of backgrounds is used.

The background process \eeto\qqbar ($n\gamma$) was generated with
{\tt PYTHIA 5.7}, while {\tt DYMU3} \cite{DYMU3} and
{\tt KORALZ} \cite{KORALZ} were used
for $\mu^+\mu^-(\gamma)$ and $\tau^+\tau^-(\gamma)$,
respectively.
The generator
of reference~\cite{BAFO} was used for \eeto\ee\ events.

Processes leading to four-fermion final states,
$(\Zn/\gamma)(\Zn/\gamma)$, 
$\Wp \Wm $, \Wev\ and \Zee,
were also generated using {\tt PYTHIA}.
The calculation of the four-fermion
background was verified using the program {\tt EXCALIBUR}
\cite{EXCALIBUR}, which consistently
takes into account all amplitudes leading to a given four-fermion
final state. 

Two-photon interactions leading to hadronic final states
were generated using {\tt TWOGAM}~\cite{TWOGAM}, separating the VDM, QPM and
QCD components.
The generators
of Berends, Daverveldt and Kleiss~\cite{BDK} were used for the leptonic
final states.

The cosmic radiation background was studied using the data collected
before the beginning of the 1998 LEP run.

The generated signal and background events were passed through the
detailed simulation~\cite{performance}
of the DELPHI detector 
and then processed
with the same reconstruction and analysis programs used for real 
data.

%
%
\section{Data selection}
\label{dataselection}
\subsection{Neutralino pair production}
\label{neutra}

\begin{figure}[htbp]\centering
\epsfxsize=16.0cm
\centerline{\epsffile{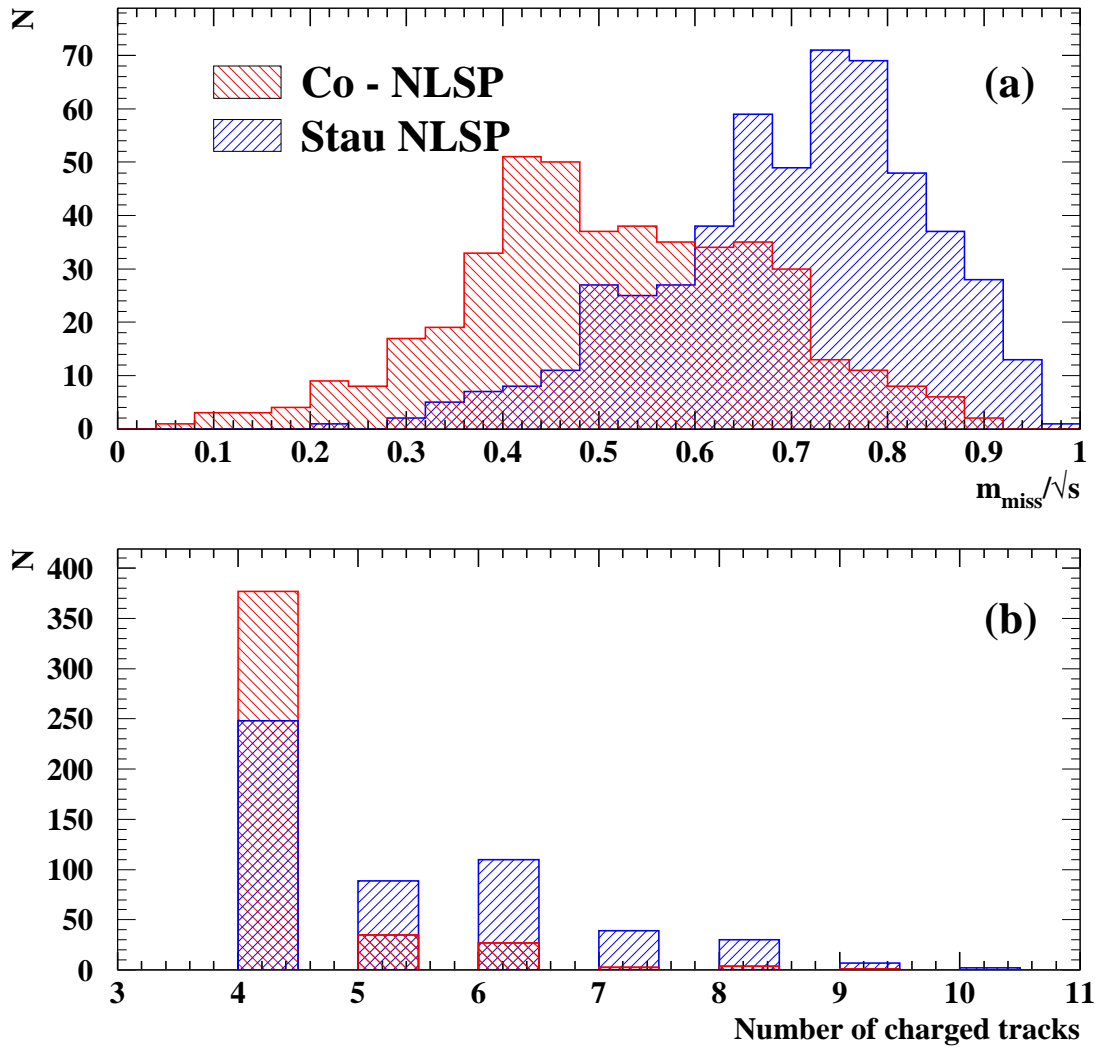}}
\caption[]{Two examples of kinematic differences between the \stuno -NLSP 
and co-NLSP scenarios. 
Fig.~(a) shows the distribution of the missing mass normalized to the 
centre-of-mass energy ($m_{miss}/\sqrt{s}$) for simulated sets with same 
mass neutralinos and same mass \stuno\ and slepton. Fig.~(b) shows the 
number of charged tracks per event for the same two sets of simulated 
signals. Histograms with positive-slope shading show a set of 
$m_{\tilde{\chi}^0_1} = 82$~GeV/c$^2$ and 
$m_{\tilde{\tau}_1} = 80$~GeV/c$^2$. Histograms with negative-slope shading 
show a set of $m_{\tilde{\chi}^0_1} = 82$~GeV/c$^2$ and 
$m_{\tilde{l}_R} = 80$~GeV/c$^2$.}
\label{fig:dif}
\end{figure}

The selection used in the search for the process
$e^+e^- \to \tilde{\chi}^0_1 \tilde{\chi}^0_1 \to 
\tilde{\tau}_1 \tau \tilde{\tau}_1 \tau  \to \tau \tilde{G} \tau \tau 
\tilde{G} \tau $\ has been described in~\cite{nuestro_papel}. A very similar 
selection was used in the general search for
$e^+e^- \to \tilde{\chi}^0_1 \tilde{\chi}^0_1 \to 
\tilde{l} l \tilde{l'} l'  \to l \tilde{G} l l' \tilde{G} l'$ 
within the co-NLSP scenario, where 
BR$(\tilde{\chi}^0_1 \to \tilde{l} l) = 1/3$\ for each leptonic flavour. 
The main two differences between these two cases
comes from the fact that the mean 
number of neutrinos carrying away undetected energy and momentum and the 
number of charged tracks per event is considerably bigger for the \stuno -NLSP 
scenario.
These 
differences can be appreciated in fig.~\ref{fig:dif}, where the simulated 
missing energy normalized to the centre-of-mass energy  and the number of 
charged tracks are represented for events with neutralinos weighing 
82~GeV/c$^2$, and staus of 80~GeV/c$^2$, compared to events with 
same-mass neutralinos and degenerate sleptons of 80~GeV/c$^2$.

The pre-selection of events is common to both scenarios, and has been 
described in~\cite{nuestro_papel}, together with the selection of the 
search for
$\tilde{\chi}^0_1 \to \tilde{\tau}_1 \tau$, which has not been changed for 
the present analysis.
Only the details of the search 
$\tilde{\chi}^0_1 \to \tilde{l} l$ are presented in the following.
Two sets of cuts were applied in order to reduce the 
$\gamma\gamma$\ and ${\rm f}{\bar{{\rm f}}}(\gamma)$\ backgrounds
and a third set of cuts to select events according to their topology:

\begin{itemize}
\item[1-] {\underline{Cuts against $\gamma\gamma$\ backgrounds}:
the transverse energy, ${\rm E_T}$, should be bigger than
4 \GeV .
The energy in a cone of $30^{\circ}$
around the beam axis was further 
restricted to be less than 
60\% of the total visible energy to avoid possible bias from the
Monte Carlo samples. The missing mass should be smaller 
than 0.88$\sqrt{s}$.
The momentum of the charged particle with largest momentum 
should 
be bigger than 8 \GeVc . The transverse missing momentum, ${\rm p_T}$,
should 
be bigger than 6 \GeVc .
These cuts reduced the $\gamma\gamma$ background  
by a factor of the order of 40.}

\item[2-] {\underline{Cuts against ${\rm f}{\bar{{\rm f}}}(\gamma)$\  
and 4-fermion backgrounds}:
the number of good tracks should be smaller than 7.
The maximum thrust was further reduced from 0.99 to 0.95. 
Dividing each event into two jets with the Durham algorithm, 
its
acoplanarity should be bigger than $8^{\circ}$. The missing mass of the 
events should be bigger 
than 0.2$\sqrt{s}$.
After these cuts, the ${\rm f}{\bar{{\rm f}}}(\gamma)$\ and 4-fermion 
backgrounds were reduced by a 
factor of the order of 30.}

\item[3-] {\underline{Cuts based on topology}:
signal events tend naturally to cluster into a 4-jet topology.  
When events are forced into a 4-jet configuration, all jets should 
be at least $18^{\circ}$\ away from the beam direction. 
When reduced by the jet algorithm into a 2-jet configuration, 
the charged particles belonging to each of these jets should be 
in a cone broader than $25^{\circ}$. Finally, the axes of
each of the four jets should be separated from the others at 
least by $9^{\circ}$.}
\end{itemize}

\begin{figure}[htbp]\centering
\epsfxsize=16.0cm
\centerline{\epsffile{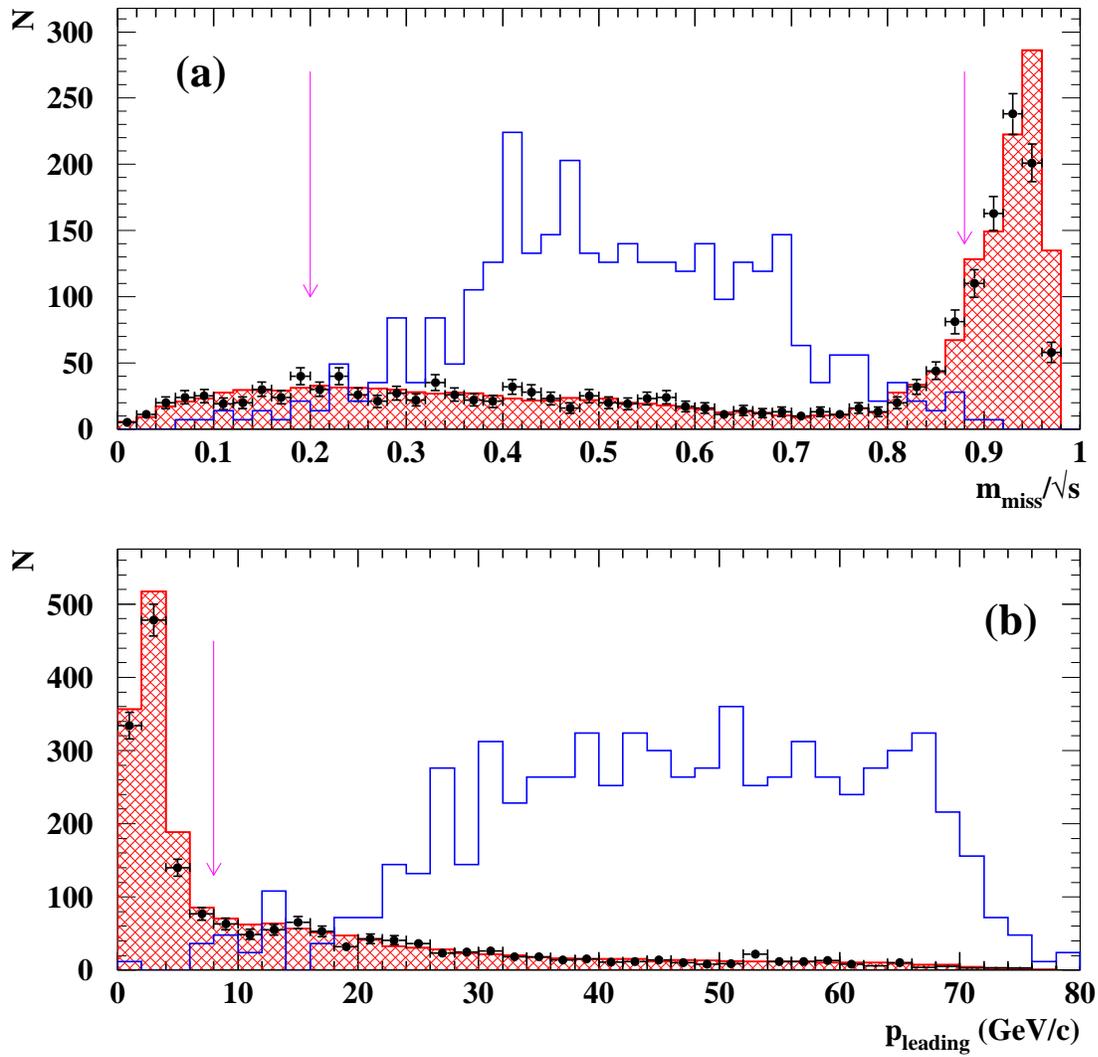}}
\caption[]{(a) Normalized missing mass 
and (b) momentum of the leading charged particle,
for data (dots), Standard Model 
simulation (cross-hatched histogram) and one of the simulated signals with 
cross-section with arbitrary normalization 
(blank histogram) after preselection. 
The arrows indicate selection criteria imposed as explained in the text.}
\label{fig:cut1b}
\end{figure}

\begin{figure}[htbp]\centering
\epsfxsize=16.0cm
\centerline{\epsffile{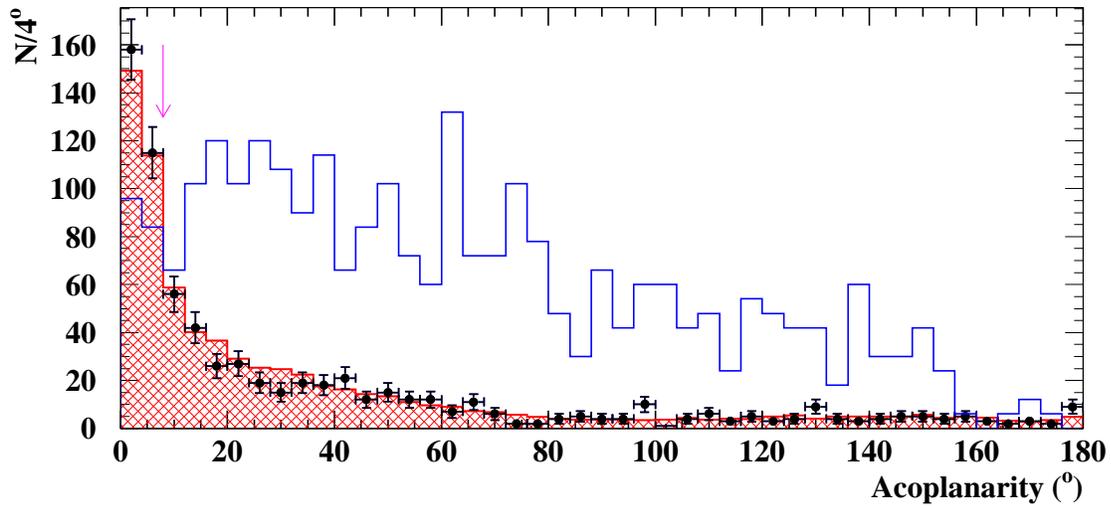}}
\vspace{-8cm}
\caption[]{ Acoplanarity of data (dots),
Standard Model background simulation 
(cross-hatched histogram) and one of the simulated signals with 
cross-section not to scale (blank histogram),
after the cut to remove 
$\gamma\gamma$\ events. 
The arrow indicates selection criterion imposed as explained in the text.}
\label{fig:cut2a}
\end{figure}

\begin{figure}[htbp]\centering
\epsfxsize=16.0cm
\centerline{\epsffile{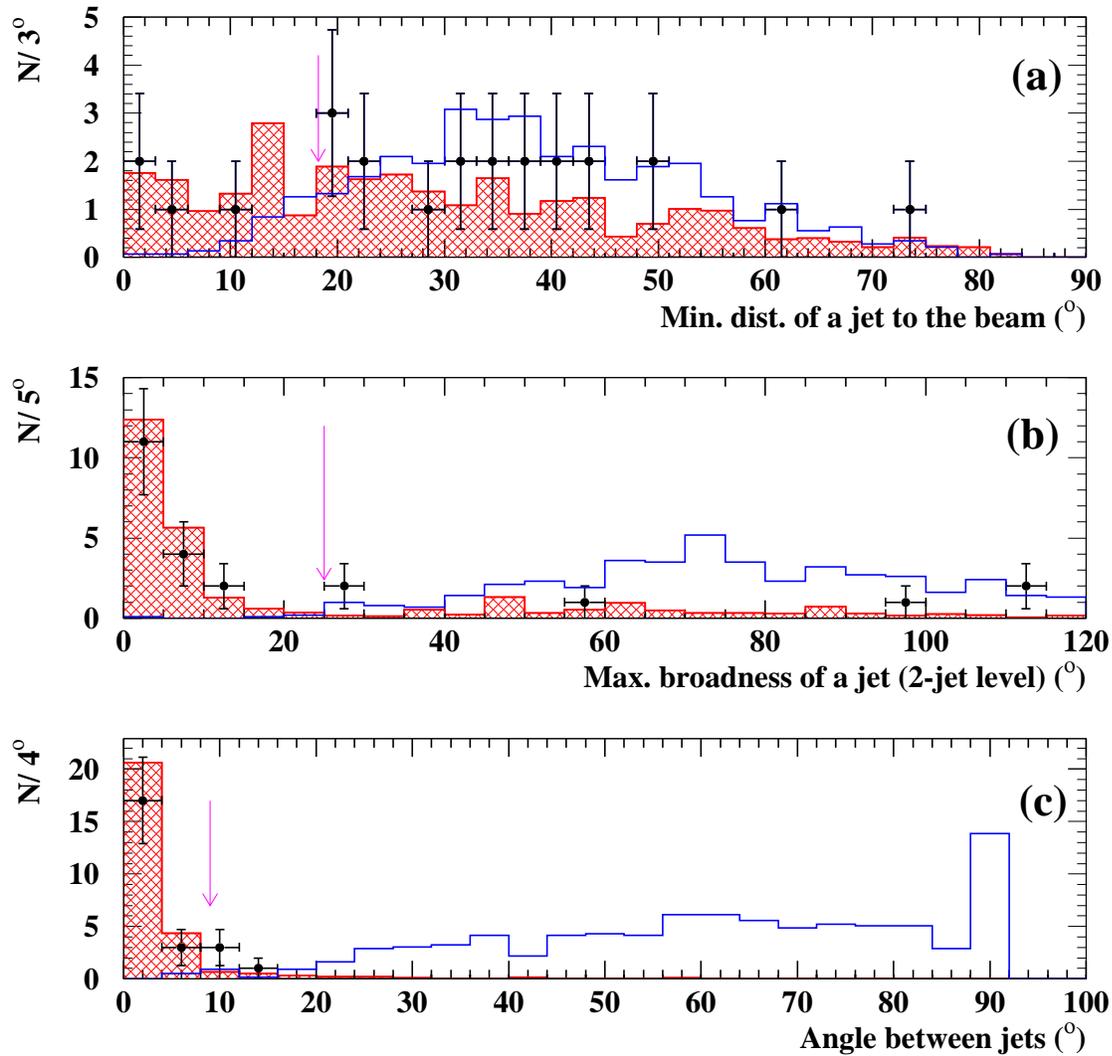}}
\caption[]{(a) Minimum angle of a jet to the beam, 
(b) maximum of angular broadness of the two jets at the 2-jet level and
(c) minimum angle between jets at the 4-jet level, 
for data (dots), Standard Model background simulation 
(cross-hatched histogram) and one of the simulated signals with 
cross-section not to scale (blank histogram),
 after the cut to remove 
$f\bar{f}(\gamma)$\ and 4-fermion events. 
The arrows indicate selection criteria imposed as explained in the text.}
\label{fig:cut3}
\end{figure}

Figures~\ref{fig:cut1b} to~\ref{fig:cut3} show some of the
distributions relevant for these selection criteria. 
The discrepancy between data and simulation 
on the last two bins of figure~\ref{fig:cut1b}-a is in a region of soft 
$\gamma\gamma$ events that is not relevant to the final results.
Tables~\ref{tab:evol} and~\ref{tab:evol-tau} show 
the effect of these cuts on the data, expected 
background and the  
signal for $m_{\tilde \chi^0_1}$ = 87 \GeVcc\ and 
$m_{\tilde \tau_1}$ =  75 \GeVcc\ for the cases 
$\tilde{\chi}^0_1 \rightarrow l \tilde{\l}$\ and
$\tilde{\chi}^0_1 \rightarrow \tau \tilde{\tau}$\ respectively. 

One event was observed to pass the search for neutralino pair production in 
the \stuno -NLSP scenario, with  $1.16 \pm 0.19$\ SM
background events expected. Two events pass the search for neutralino pair 
production in 
the co-NLSP scenario, with  $1.2 \pm 0.30$\ SM
background events expected.
After these cuts, efficiencies between 27.0 and 40.7\% were obtained for 
the signal events.

\subsection{Slepton pair production}
\label{stau}
This section describes the update of the search 
for the process
$e^+ e^- \to \tilde{\tau}_1 \tilde{\tau}_1 \to \tau \tilde{G} \tau \tilde{G}$
already described in~\cite{nuestro_papel,ref:flying}.
An additional 153.6~${\mathrm pb}^{-1}$ integrated luminosity 
collected at the centre-of-mass energy
of 189 GeV has been analysed using the same procedure as for the data
collected at 183 GeV and using the same values for the data selection cuts.
The same selection cuts have been applied to the search for
$\tilde{e}_R$-- and $\tilde{\mu}_R$- pair production in the framework of
$\tilde{l}$ co-NLSP scenario. 
Therefore, only results and efficiencies will be reported in this section,
since the details of the selection criteria can be found
in~\cite{nuestro_papel,ref:flying}.  

\subsubsection{Search for secondary vertices}
\label{searchkinks}
\hspace{\parindent}

This analysis exploits a peculiarity of the
$\tilde{l} \rightarrow l \tilde{G}$ 
topology in the 
case of intermediate gravitino masses 
(i.e. 0.5\eVcc\ $\lesssim m_{\tilde{G}}\lesssim$\ 200\eVcc\ as dictated by 
eq.~\ref{life}), namely,
one or two tracks coming from the interaction point and at least one
of them with either a secondary
vertex or a kink. 

Rather loose preselection cuts
were imposed on the events in order to suppress the low energy background
(beam-gas, beam-wall, etc), $\gamma \gamma$, $e^+e^-$ and hadronic events.
The only cut that was changed with respect to the analysis at 183 GeV is the
total electromagnetic energy required in the event. 
It was increased to $\sqrt{s}/2$ in order 
to improve the efficiency for selectrons.
This did not
increase noticeably the background contamination by Bhabha events.
These preselection cuts left about 0.6\% of the whole data sample.
The events that survived the preselection cuts underwent the search for
secondary vertices or kinks. 

%

\begin{figure}[hbpt]
\vspace{-1cm}
\centerline{\epsfxsize=16.0cm \epsfysize=21.0cm \epsfbox{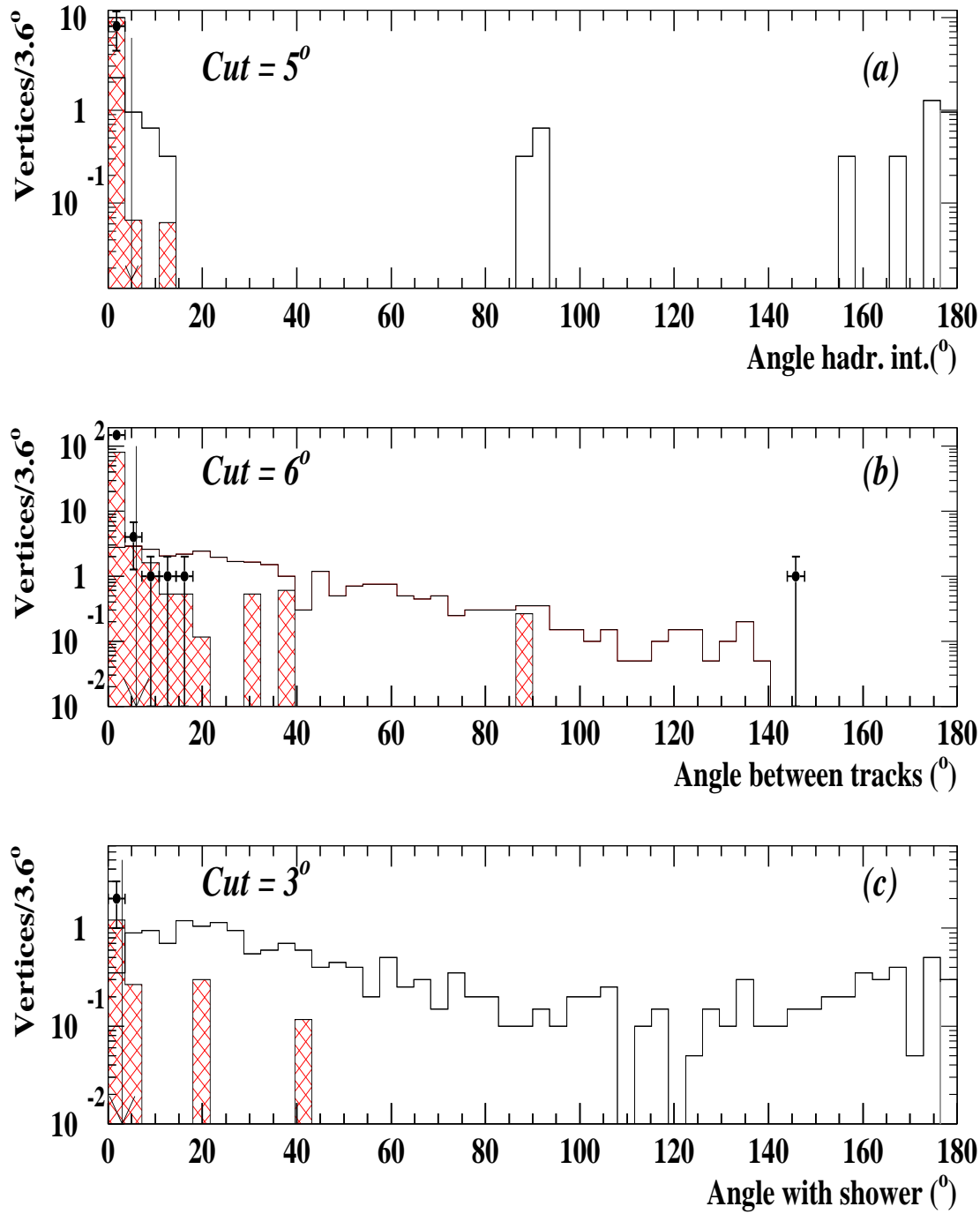}} 
   \caption[]
{(a) Angle between the directions defined by the hadronic vertex and the
   reconstructed vertex, (b) angle between the tracks of the kink, and (c)
angle between the electromagnetic shower
  and the direction defined by the difference between the momenta of the
   $\tilde{\tau}_1$ and its associated $\tau$, defined at the crossing
  point
 for real data (dots), expected Standard Model background (cross-hatched 
histogram) and simulated signal for $m_{\tilde{\tau}_1} = 60$\GeVcc\
decaying with a mean 
 distance of 50~cm (blank histogram). Events that do not have hadronic 
interactions are not included in (a), and events without electromagnetic 
showers are not included in (b).
The arrows indicate the selection criteria imposed.}
  \label{fig:grav:kinks_BG}
\end{figure}

Fake decay vertices could be present amongst the reconstructed secondary 
vertices, being produced by particles interacting in 
the detector material or by radiated photons if the particle trajectory
was reconstructed into two separated tracks.
To eliminate these classes of events, additional requirements were imposed:
\begin{itemize}
  \item[-] to reject hadronic interactions,
        any reconstructed hadronic interaction (secondary vertices 
        reconstructed in region where there is material) 
        must be outside a cone of half angle 5$^\circ$ around the slepton 
        direction; 
  \item[-] to reject segmented tracks,
        the angle between the tracks used to define a vertex
        had to be larger than 6$^\circ$;
  \item[-] to reject photon radiation
        in the case of $\tau$ clusters with only one track,
        there had to be no neutral particle  in a 3$^\circ$ cone
        around the direction defined by the difference between the
        $\tilde{\tau}_1$ momentum and the momentum of the
        $\tau$ daughter calculated at the crossing point.
\end{itemize}
If no pair of tracks was found to survive these conditions, the
event was rejected.
Fi\-gu\-re~\ref{fig:grav:kinks_BG} shows the distribution of these three 
quantities.
The distributions compare 
real data, expected Standard Model background 
simulation and simulated signal 
for $m_{\tilde{\tau}_1} = 60$\GeVcc\ decaying with a mean decay length of
50~cm. The excess of data in the first bins of fig.~\ref{fig:grav:kinks_BG}-b 
is due to an underestimation in the simulation of mismatchings between the 
tracking devices.
%
%
%

One event in real data was found to satisfy all the conditions described
above, while $1.18^{+0.63}_{-0.35}$\ were expected from SM backgrounds. 
The event was compatible with a $\gamma \gamma \rightarrow 
\tau^+\tau^-$ with a hadronic interaction in the ID detector.


The vertex reconstruction procedure was sensitive to radial decay lengths, R, 
between 20~cm and 90~cm. Within this region a vertex was 
reconstructed with an efficiency of $\sim$52\%. The VD 
(Vertex Detector) and the ID (Inner Detector)  
were needed to reconstruct the
$\tilde{\tau}_1$\ track and the TPC (Time Projection Chamber)
to reconstruct the decay products. 
The shape of the efficiency distribution was essentially flat as a 
function of R, going down when
the \stuno\ decayed near the outer surface of the TPC, due to inefficiencies 
in the reconstruction of the tracks coming from the desintegration products of 
the $\tau$. 
Also, 
the sensitive region and the efficiency of the vertex reconstruction at 189
GeV was slightly lower than at 183\GeV\  due to the loss of tracks 
not reconstructed when their VD hits had no information in the 
{\it z} direction. Such tracks were not reconstructed for the 189\GeV\ run.
However, some of the efficiency lost in the vertex search was recovered later 
by the search based on  large impact parameter.

The search for vertices had 
an efficiency of the order of 46\% for \stuno\ masses between 40 and 
85\GeVcc\ with a mean decay length of 50~cm. The efficiencies decreased 
near the kinematical limit due to a small boost that allowed for big angles to 
appear between the \stuno\ and the desintegration products of the $\tau$.
For \stuno\ masses below 40\GeVcc, the efficiency decreased gradually 
due to the cut that rejects segmented tracks. This happened 
because the resulting big boost causes the angle between \stuno\ and 
$\tau$\ decay products to be very small. 

As already said, the same selection criteria was applied to smuons and
selectrons. The 
efficiency for selectrons decreased due to the preselection cut on total 
electromagnetic energy (lower than $0.5\sqrt{s}$) and was around 31\% for 
$m_{\tilde{e}_R}$ between 40 to 85\GeVcc, while the smuon efficiency 
increased to 55\% for the same mass range.

\subsubsection{Large impact parameter search}
\label{searchimpact}
\hspace{\parindent}

To investigate a region of lower gravitino masses 
the previous search was extended to the case of sleptons with mean decay length
between 0.25 cm and approximately 10 cm. In this case the \slep~
track is not reconstructed and only the $l$
(or the decay products in the case of \stau) is detected.
The impact parameter search was only applied to those events accepted by the
same general requirements as in the search for secondary vertices, and not 
selected by the vertex analysis. 
The same selection criteria described in
references~\cite{nuestro_papel,ref:flying} were applied.

The efficiencies were derived for the different $\tilde{\tau}_1$ masses and
decay lengths by applying the same selection to
the simulated signal events.
The maximum efficiency was 32\% corresponding to a mean decay
length of 2.5~cm. The efficiency decreased very fast for lower decay lengths
due to the cut on minimum impact parameter. 
 The efficiency at 189 GeV was slightly larger than at 183\GeV\  
since some events not passing the secondary vertex selection were recovered
in this search, as explained before. For longer decay lengths, 
the appearance of reconstructed $\tilde{\tau}_1$ tracks in combination with
the cut on the maximum amount of charged particle tracks caused the
efficiency to decrease smoothly. This decrease is compensated by a rising
efficiency in the search for secondary vertices.
For masses above 30\GeVcc\ no dependence on the  $\tilde{\tau}_1$ mass was 
found far from the kinematic limit. However for lower masses the efficiency
decreased and it was almost zero for a 5\GeVcc\ \stuno.

The same selection was applied to selectrons and smuons. For smuons
the efficiency increased to 59\% for a mean decay 
length of 2.5 cm and masses over 
30\GeVcc\ since the smuon has always one prong decay. For selectrons the 
efficiencies were almost the same as those for staus. 

Trigger efficiencies were studied simulating the DELPHI trigger
response to the events selected by the vertex search and by the large impact
parameter analysis, and were found to be around 99\%.

No events in the real data sample were selected with the above criteria, while
$0.32^{+0.19}_{-0.10}$\ were expected from SM backgrounds.
The number of expected background events at $\sqrt{s}=189$~GeV is shown in
Table~\ref{tab:grav:bkg} for the combination of the vertex and large
impact parameter searches.

\noindent

\subsubsection{Small impact parameter search}
\label{smallimpact}
\hspace{\parindent}

The large impact parameter search can be extended further down to 
mean decay lengths of around 0.1~cm. The same selection criteria described in 
reference~\cite{nuestro_papel} was applied. However, some extra selection was
added in order to 
reduce background from detector noise or failure, cosmic radiation and
$\tau \tau$ events.

Events with anomalous noise in the TPC were rejected requiring less than 20
charged particles (before track selection) and relative error of the 
measured momentum of the leading tracks (charged particles with largest 
momentum in each hemisphere) less than 50\%. The cosmic muon rejection was
improved by requiring that the leading tracks with impact parameters 
larger than
1~cm must be reconstructed in the TPC. To reduce the $\tau \tau$ background 
where one of the taus decays into a three prong topology  
(when the single track is not reconstructed ), and also
gamma conversions, any leading track must have at least other charged particle
at an angular distance larger than $5^{\circ}$.     

The efficiency of the search turned out not to depend  
on the $\tilde{\tau}_1$ mass for masses over 40 \GeVcc, but rather on 
the $\tilde{\tau}_1$ decay length in the laboratory system. 
The maximum efficiency was $\sim$ 38\% for a mean decay length of 
$\sim$ 2~cm, the efficiency dropped 
at small decay lengths ($\sim$ 15\% at 1~mm).

The same selection criteria were used to search for smuons. To search for
selectrons, in order to increase efficiency, the cut 
$(E_1+E_2)<0.7 \mathrm{E}_{beam}$\ 
(where $E_1$, $E_2$ are the electromagnetic energy deposits associated to the
leading tracks) was not applied. The Bhabha events that survived the
selection, when the previous rejection cut was not applied, were those where at
least one of the electrons underwent a secondary interaction, thus acquiring
a large impact parameter. However, it was found that in these cases the
measured momentum of the electron was smaller than the electromagnetic energy
deposition around the electron track. Therefore, the cut 
 $(E_1/p_1+E_2/p_2)<2.2$ was used for the selectron search.
The maximum efficiency reach for the smuon search was 43\% and for the 
selectron search 35\% at  2~cm mean decay length.

The number of events selected in data was 4, and 
4.54$^{+1.12}_{-0.57}$ events were expected 
from Standard Model background (see table~\ref{tab:cavallo1}). Two of the
candidates had tracks with fitting problems and the other two events were 
compatible with Standard Model $\tau \tau$ events.

\subsection{Chargino pair production}
\label{chargino}

\begin{center}
\begin{figure}[htbp]
\epsfig{figure=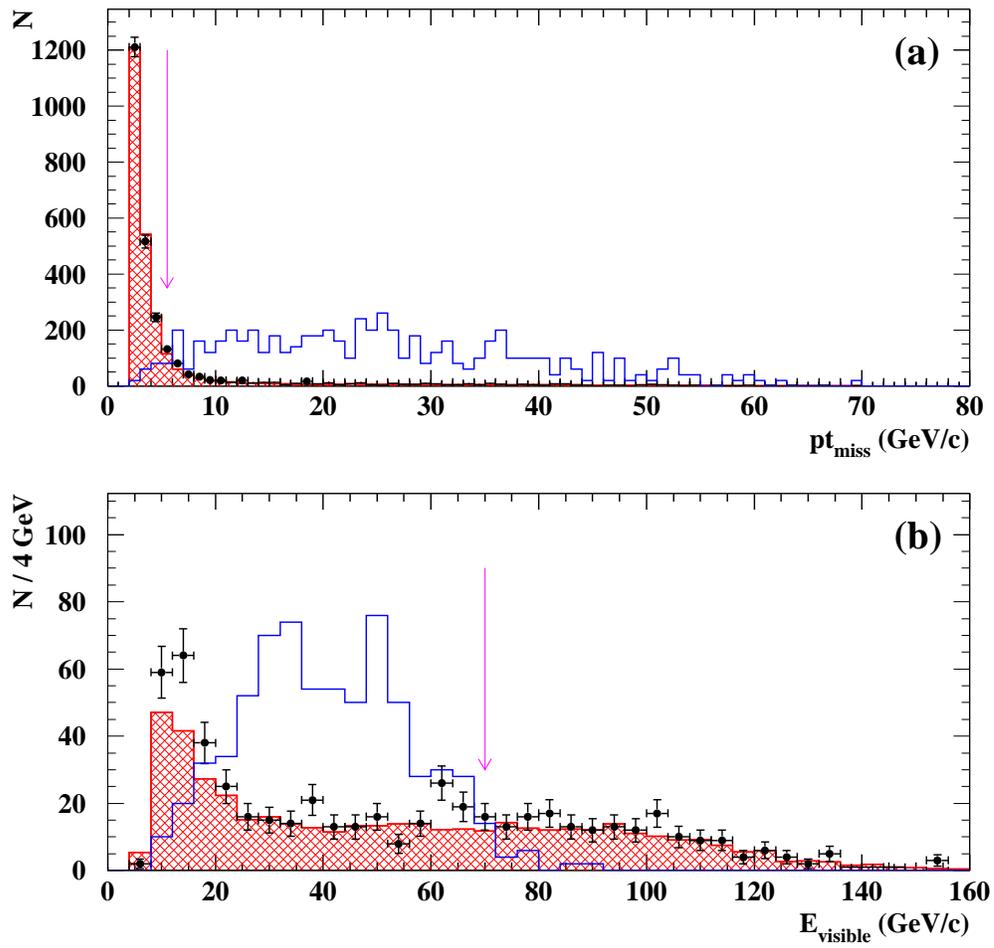,width=0.9\textwidth}
   \caption[]{Missing transverse momentum (a) and 
visible energy (b), for real data (dots), expected Standard Model 
background (shaded 
histogram) and simulated signal for $m_{\tilde{\chi}_1^+}=85$\GeVcc\ 
 and $m_{\tilde{\tau}_1^+}=69$\GeVcc\
decaying with a mean 
 distance of 50~cm (blank histogram). 
The arrows indicate selection criteria 
imposed as explained in~\cite{chargino_183}.}
\label{fig:mcdata_mssm}
\end{figure}
\end{center}

The analyses used to search for the lightest charginos varies according 
to the stau lifetime. Figures~\ref{fig:grav:kinks_BG}
and~\ref{fig:mcdata_mssm}
illustrate the 
distributions of some of the
main variables 
used in the analyses described respectively in section~\ref{stau} (the stau 
decays with big impact parameter or producing a kink),
and~\cite{chargino_183} (the stau decays at the main vertex). The plots show 
real data, expected 
Standard Model background, and a simulated signal of 
$m_{\tilde{\chi}_1^+}=85$\GeVcc\ and $m_{\tilde{\tau}_1^+}=69$\GeVcc.
For the 
three analyses described in section~\ref{experimentalprocedure}, 
Table~\ref{tab:results} shows the range of efficiencies, the 
main components and the total amount of the expected 
background events, and 
the number of observed data events for each sample.


\section{Results and interpretation}
\label{sec:resultados}

Since no evidence for a signal was found in the data, 
limits on the cross-section of sparticle pair production were derived.
In what follows, the model described in reference~\cite{Dutta} will
be used in order to derive limits. This is a model which 
assumes radiatively broken electroweak symmetry and 
null trilinear couplings at the messenger scale.
The SUSY soft parametes, gauge and Yukawa couplings are 
evolved between the electroweak scale (chosen to be $m_t$) 
and the messenger scale
following the prescription of~\cite{barger}. The masses of gauginos and 
sfermions at the messenger scale are calculated taking into account 
corrections arising from threshold effects.
The 
corresponding parameter space was scanned as follows:
$1\leq n \leq 4$, $5\ {\rm TeV}\leq\Lambda\leq 90\ {\rm TeV}$, 
$1.1\leq M/\Lambda \leq 10^9$, $1.1\leq \tan\beta\leq 50$, and 
$sign(\mu) = \pm 1 $,
where
$n$\ is the number of messenger generations in the model, $\Lambda$\ 
is the ratio between the vacuum expectation values 
of the auxiliary component and the scalar component of the
superfield and $M$\ is the messenger mass scale. The parameters 
tan $\beta$\ and $\mu$\ are defined as for MSUGRA.

\subsection{Neutralino pair production}

Limits on the cross-section for neutralino pair production were derived in 
the two scenarios for
each ($m_{\tilde{\chi}_1^0}$,$m_{\tilde{l}}$) combination. For the 
\stuno -NLSP case, the combination took also into account 
the results from the LEP
runs of 1996 and 1997~\cite{nuestro_papel}. 

\begin{figure}[htbp]\centering
\epsfxsize=14.5cm
\centerline{\epsffile{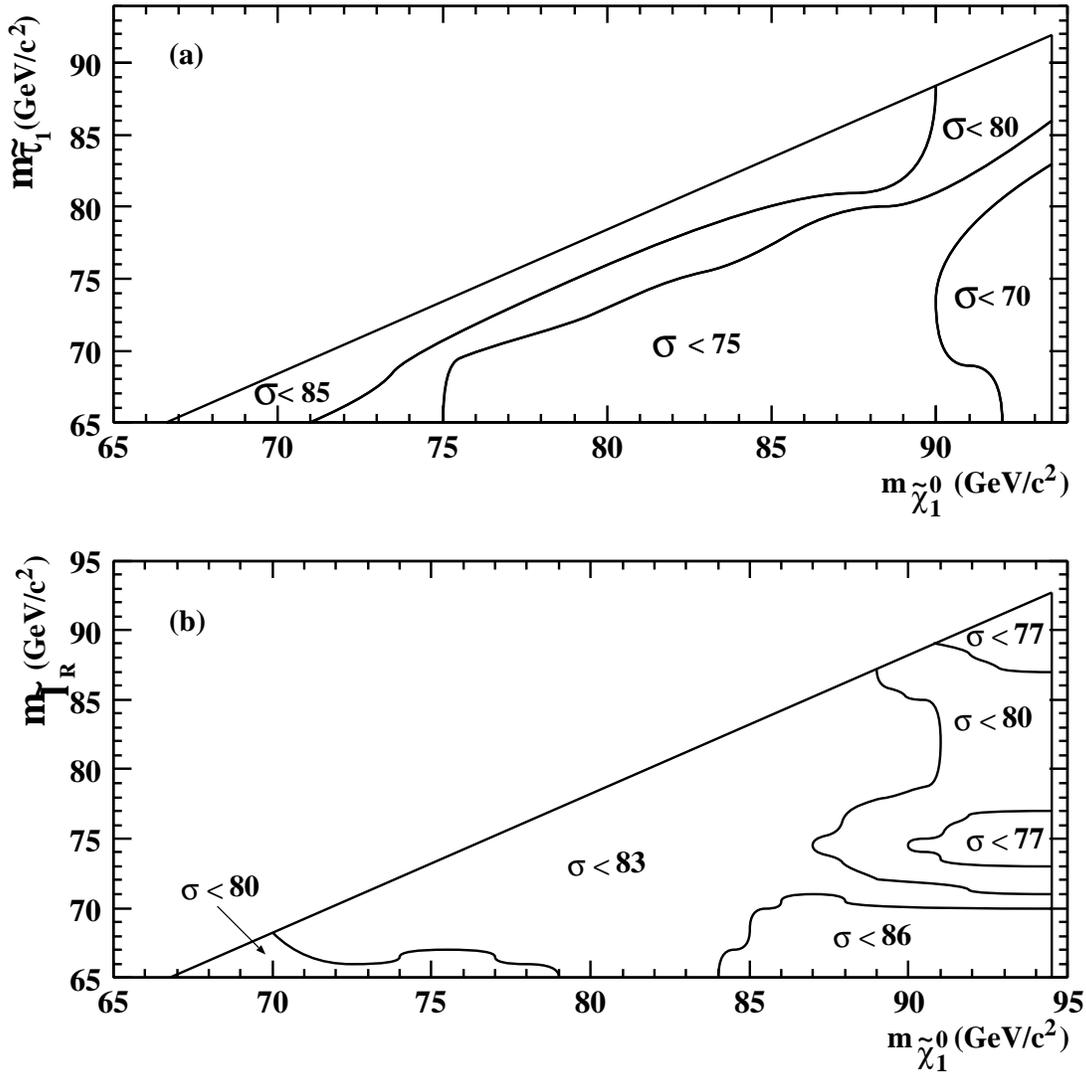}}
\caption[]{95\% C.L. upper limit of the \nuno\ pair 
production cross-section (in femtobarn) at $\sqrt{s}=18$9 \GeV~ (a)
after combining the results of the searches from $\sqrt{s} = 161$\ up to 
189 \GeV, as a
function of $m_{\tilde{\chi}_1^0}$\ and $m_{\tilde{\tau}_1}$
for the case $n=3$ in the \stuno -NLSP scenario, where $n$~ is
the number of messenger generations and (b) using data at $\sqrt{s} = 189$~GeV,
as a
function of $m_{\tilde{\chi}_1^0}$\ and $m_{\tilde{l}_R}$ in the co-NLSP 
scenario. 
The diagonal and vertical lines
show respectively the limits $m_{\tilde{\chi}_1^0} = 
m_{\tilde{\tau}} + m_{\tau}$
and $m_{\tilde{\chi}_1^0} = \sqrt{s}/2$.
}
\label{fig:limit}
\end{figure}

Figure~\ref{fig:limit}-a 
shows the 95\% C.L. upper limit for the \nuno\ pair 
production cross-section at $\sqrt{s} = 189$ GeV as a
function of $m_{\tilde{\chi}_1^0}$\ and $m_{\tilde{\tau}_1}$\ 
after combining the results of the searches from $\sqrt{s}$ = 161 up to 
189 GeV with the likelihood ratio method~\cite{Read}, and scanning through 
the whole parameter space. 
Figure~\ref{fig:limit}-b 
shows the 95\% C.L. upper limit for the \nuno\ pair 
production cross-section at $\sqrt{s}$ = 189 GeV as a
function of $m_{\tilde{\chi}_1^0}$\ and $m_{\tilde{l}_R}$.
For different number of messenger generations,
the ratios between production cross sections at different energies
are bound to  vary within certain limits. Figure~\ref{fig:limit} presents 
as an example the case of $n=3$. For the other scenarios considered 
in this study ($n = 1$, 2 and 4), the maximum difference with respect to 
figure~\ref{fig:limit} is not bigger than 10\%. This variation 

\subsection{Slepton pair production}

\begin{figure}[htbp]\centering
\vspace{-2.cm}
\centerline{\epsfxsize=16.0cm \epsfysize=20.0cm \epsffile{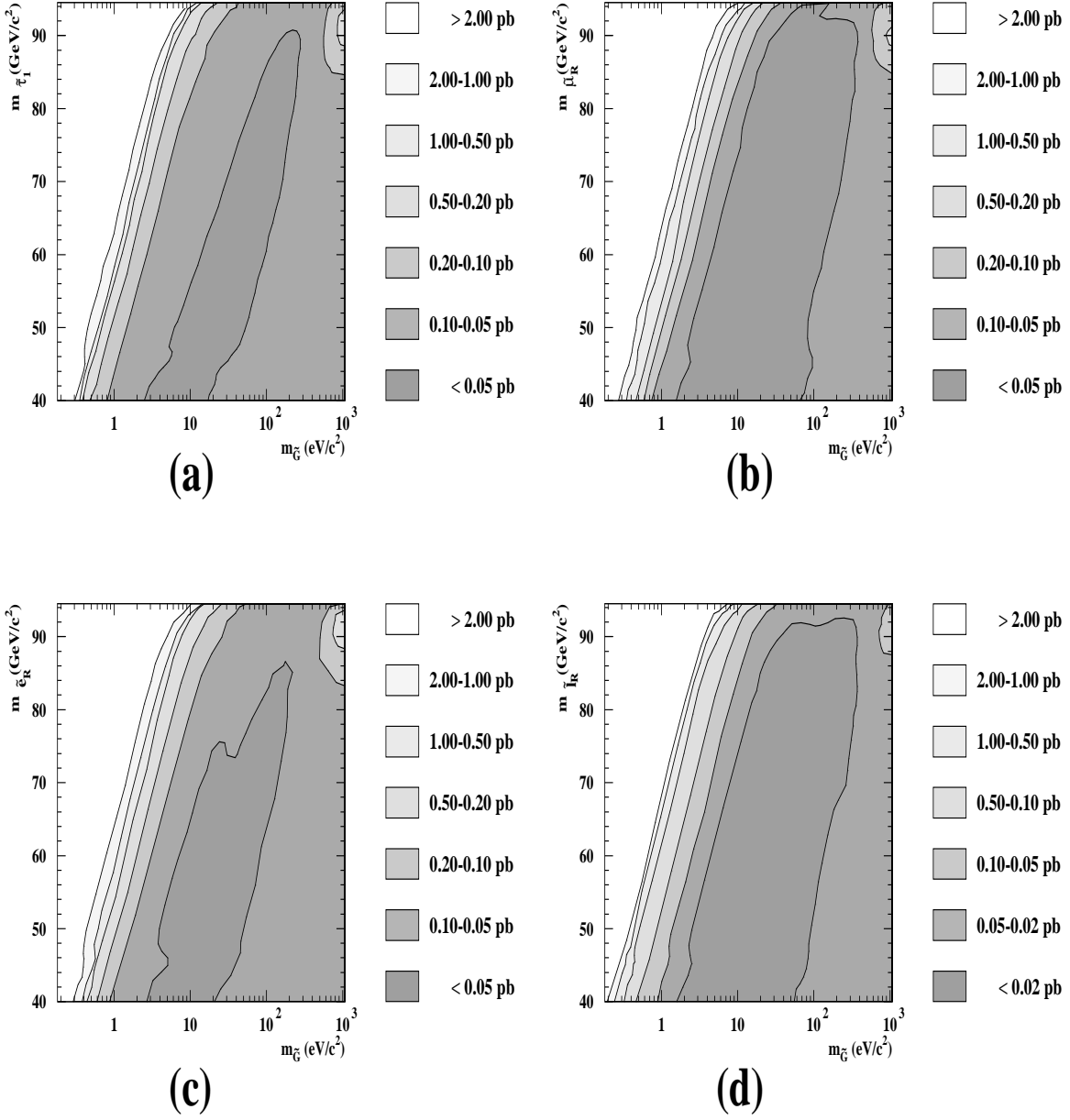}}
  \caption[]{
95\% C.L. upper limit of the $e^+e^-\to\tilde{\tau}_1\tilde{\tau}_1$\  (a),
$e^+e^-\to\tilde{\mu}_R\tilde{\mu}_R$\  (b),
$e^+e^-\to\tilde{e}_R\tilde{e}_R$\  (c), and
$e^+e^-\to\tilde{l}_R\tilde{l}_R$\  (d)   
production cross-sections,  at $\sqrt{s}$=189\GeV\ 
after combining the results of the searches at $\sqrt{s} = 130-$189\GeV. 
Results are shown in the 
($m_{\tilde{G}}$,$m_{\tilde{l}}$) plane. Searches for events containing
charged particle tracks  with small impact parameter, large impact 
    parameter, secondary vertices and the search for heavy stable leptons are
    combined.}  
  \label{fig:cross_todos}
\end{figure}

Figure~\ref{fig:cross_todos} 
shows the 95\% C.L. upper limit on the slepton pair production
cross-section  at $\sqrt{s} = 1$89~GeV
after combining the results of the searches at $\sqrt{s}$ = 130-189 GeV 
with the likelihood ratio method~\cite{Read}.
The results are presented in the 
($m_{\tilde{G}}$,$m_{\tilde{l}}$) plane combining the two impact
parameter searches, the secondary vertex analysis and the stable heavy lepton
search~\cite{heavyparticles}. 
In particular, figure~\ref{fig:cross_todos}-a shows that
the minimum upper limits achieved for a given $\tilde{\tau}_1$ were around
0.05-0.10~pb depending on $m_{\tilde{G}}$.
For $m_{\tilde{G}} > 9~{\mathrm eV}/c^2$ and a 
80\GeVcc\ $\tilde{\tau}_1$, a 0.10~pb limit was obtained. 
Figures~\ref{fig:cross_todos}-b and -c show the corresponding upper limit
for $\tilde{\mu}_R$- and $\tilde{e}_R$ pair production cross-sections. 
Assuming mass degeneracy 
of the three supersymmetric particles, $\tilde{\tau}_1$, $\tilde{e}_R$ and
$\tilde{\mu}_R$, figure \ref{fig:cross_todos}-d shows the 95\%
C.L. upper limit for the $\tilde{l}_R$ pair production cross-section.


\subsection{Chargino pair production}

\begin{figure}[htbp]\centering
\epsfxsize=16.0cm
\centerline{\epsffile{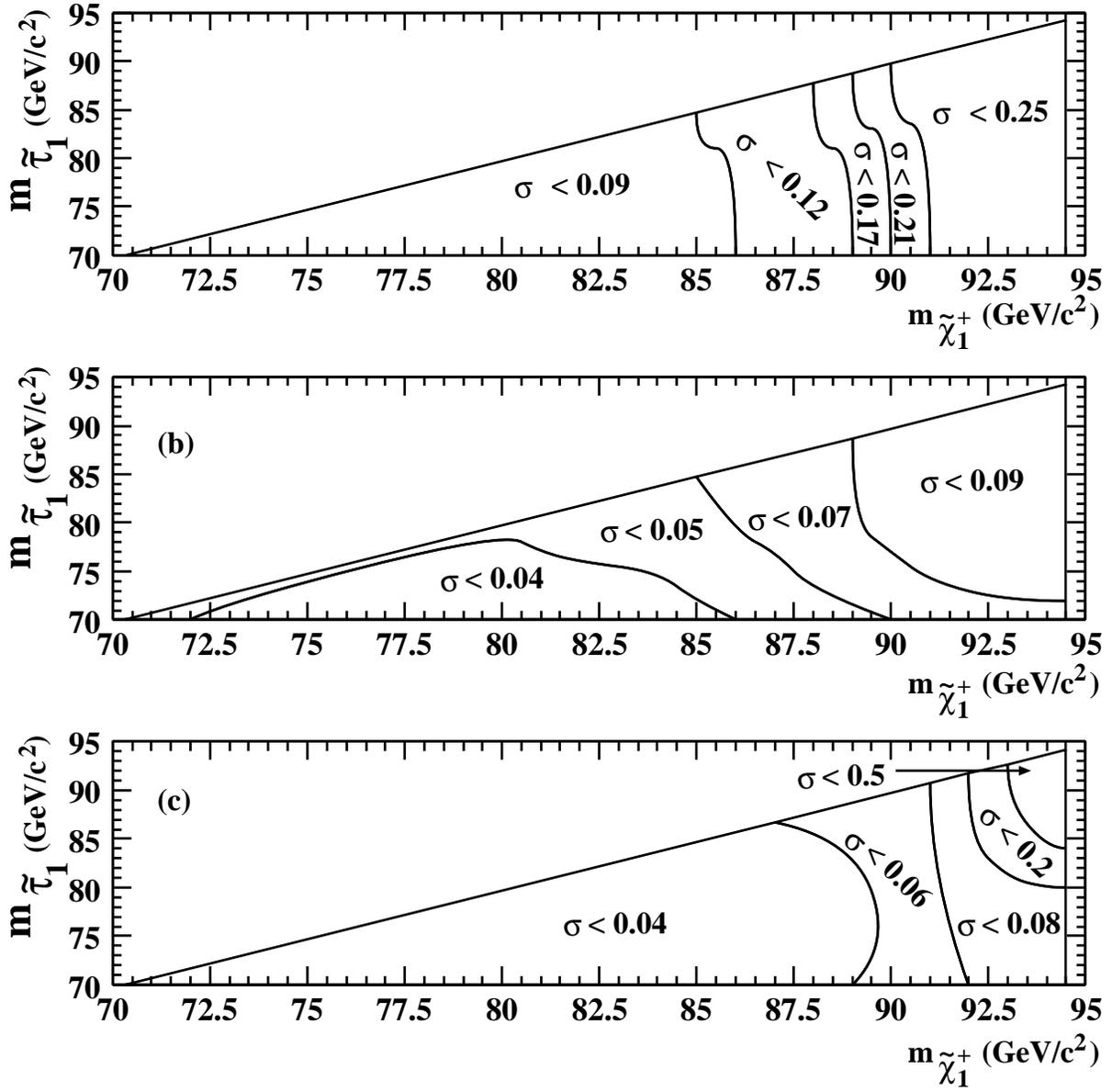}}
\caption{Limits in picobarn on the lightest chargino pair production 
cross-section at 
95\% CL. Limits are shown as functions of $m_{\tilde{\chi}_1^+}$\ and 
$m_{\tilde{\tau}_1}$ for (a) $m_{\tilde{G}}=1$\eVcc , 
(b) $m_{\tilde{G}}=1$00\eVcc\ and (c)
 $m_{\tilde{G}}=1$000\eVcc .}
\label{fig:xsec}
\end{figure}

Limits on the production cross-section for chargino pairs were derived for
each ($m_{\tilde{G}}$,$m_{\tilde{\tau}_1}$,$m_{\tilde{\chi}_1^+}$) 
combination.
Figure~\ref{fig:xsec} 
shows the 95\% C.L. upper limit on the chargino pair 
production cross-section at $\sqrt{s} = 189$~GeV as a
function of $m_{\tilde{\chi}_1^+}$\ and $m_{\tilde{\tau}_1}$\ 
after combining the results of the searches at $\sqrt{s} = 18$3 
and 189 \GeV\ with the maximum likelihood ratio method~\cite{Read} for 
$\Delta m \geq 0.3$\ \GeVcc\ and
$m_{\tilde{G}}=1$, 100 and 1000\eVcc . 
These
limits, which directly 
reflect the efficiencies of the
applied selections, can be understood as follows:

\begin{description}
\item[$m_{\tilde{G}}=1$\eVcc :] 
The efficiency of this analysis depends mainly on the mass of the 
chargino. To smaller chargino masses correspond bigger event missing energies,
and bigger efficiencies.

\item[$m_{\tilde{G}}=100$\eVcc :] The map of efficiencies is the result of 
the convolution of two factors. First, 
larger stau masses imply a 
smaller lifetime, and hence a smaller efficiency. Second, 
a larger chargino mass leads
to smaller stau momenta, and to smaller 
decay lengths.

\item[$m_{\tilde{G}}=1000$\eVcc :] In this case, the map of efficiencies 
is mainly 
affected by the momentum of the stau, because the method used to identify 
heavy stable particles relies on the lack of Cherenkov radiation 
in DELPHI's RICH detectors. 
To remove SM backgrounds, low momentum particles 
are removed, thus reducing the efficiency for higher chargino masses, 
especially in the region of small  $\Delta m$.
\end{description}

\subsection{Interpretation}

\subsubsection{Neutralino pair production}
\begin{figure}[htbp]\centering
\centerline{\epsfxsize=14.0cm \epsffile{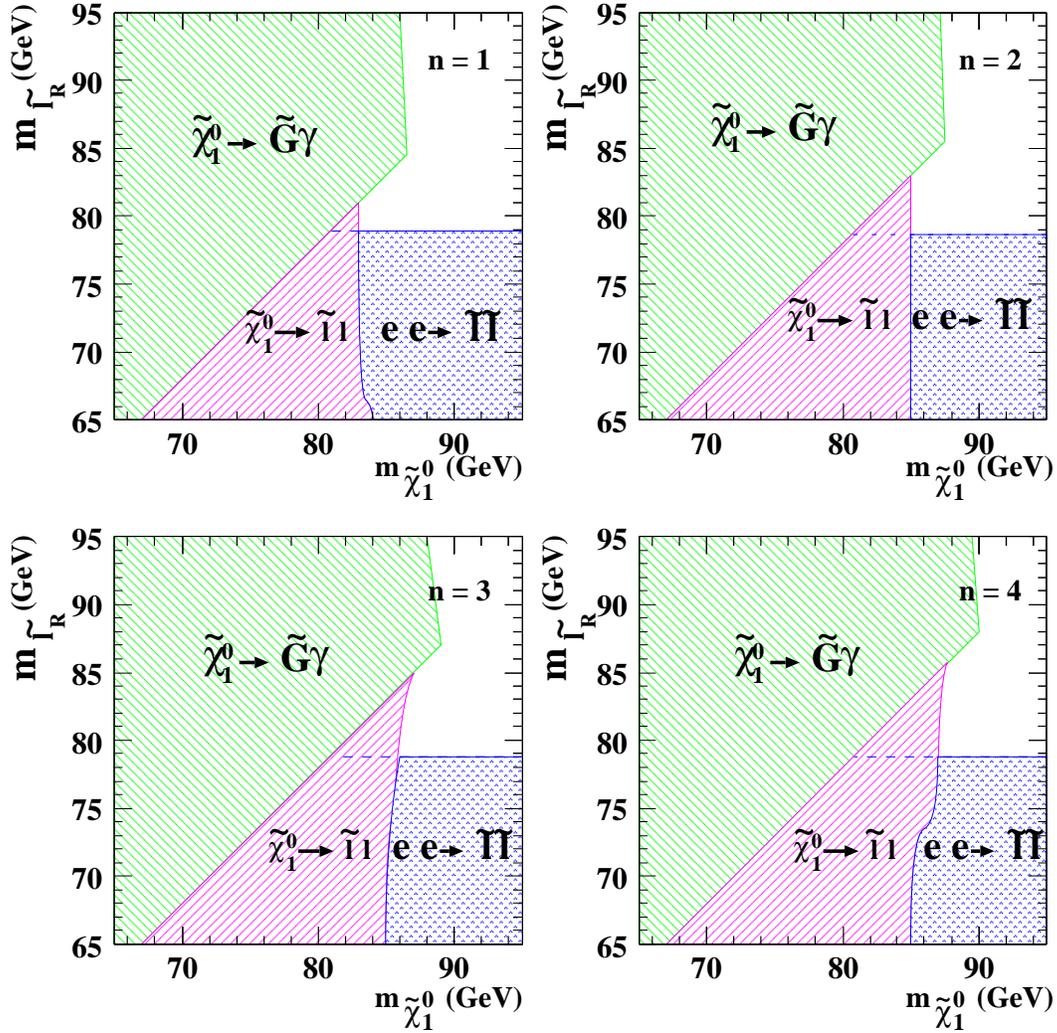}}
\vspace{-0.5cm}
\caption[]{Areas excluded at 95\% C.L. with $m_{\tilde{G}}< 1$ \eVcc~ in the 
$m_{\tilde{\chi}_1^0}$\ {\it vs.} $m_{\tilde{l}_R}$ plane for $n = $1 to 4. 
The positive-slope dashed area is excluded by this analysis.
The negative-slope dashed
area is excluded by the search for
$\tilde{\chi}^0_1\rightarrow \gamma \tilde{G}$,
 and the point-hatched area by the direct search for slepton pair 
production in the 
MSUGRA framework.}
\label{fig:masses}
\end{figure}

Given the aforementioned limits for the production cross-section, 
some sectors of the 
($m_{\tilde{\chi}^0_1},m_{\tilde{l}}$) space can be excluded.
In order to achieve the maximum sensitivity, the results from 
two other analyses are taken into account. 
The first is the search for slepton pair production
in the context of gravity mediated SUSY breaking models~\cite{slep_189}. 
In the case where the MSUGRA $\tilde{\chi}_1^0$\ is 
massless, the kinematics corresponds to the case of $\tilde{l}$\ 
decaying into a lepton and a gravitino.
The second is the search for lightest neutralino pair production 
in the region of the mass space where  
$\tilde{\chi}_1^0$\ is the NLSP~\cite{2gamma_189} 
(the region above the diagonal 
line in fig.~\ref{fig:masses}, i.e. $m_{\tilde{\tau}} > m_{\tilde{\chi}^0_1}$). 
Within this zone, the neutralino decays into a gravitino and a photon.

As an illustration, fig.~\ref{fig:masses} 
presents the 95\% C.L. excluded areas for $m_{\tilde{G}} < 1$\ \eVcc\ in the 
$m_{\tilde{\chi}_1^0}$\ {\it vs.} $m_{\tilde{l}_R}$ plane for the 
co-NLSP case. 
The positive-slope dashed area 
is excluded by this analysis. The resulting 95\% C.L. 
lower limit for the mass of the lightest neutralino is 82.5 \GeVcc .
The negative-slope dashed area is excluded by
the analysis searching for neutralino pair production followed by the decay
$\tilde{\chi}^0_1\rightarrow \tilde{G}\gamma$. 
The point-hatched
 area is excluded by the direct search for slepton pair
production within MSUGRA scenarios~\cite{slep_189}.

Table~\ref{tab:mass} shows the 95~\% C.L. lower limits on the mass of the 
lightest neutralino within
the two scenarios for different $n$.

\subsubsection{Slepton pair production}

\begin{figure}[htbp]
\centerline{\epsfxsize=16.0cm \epsfbox{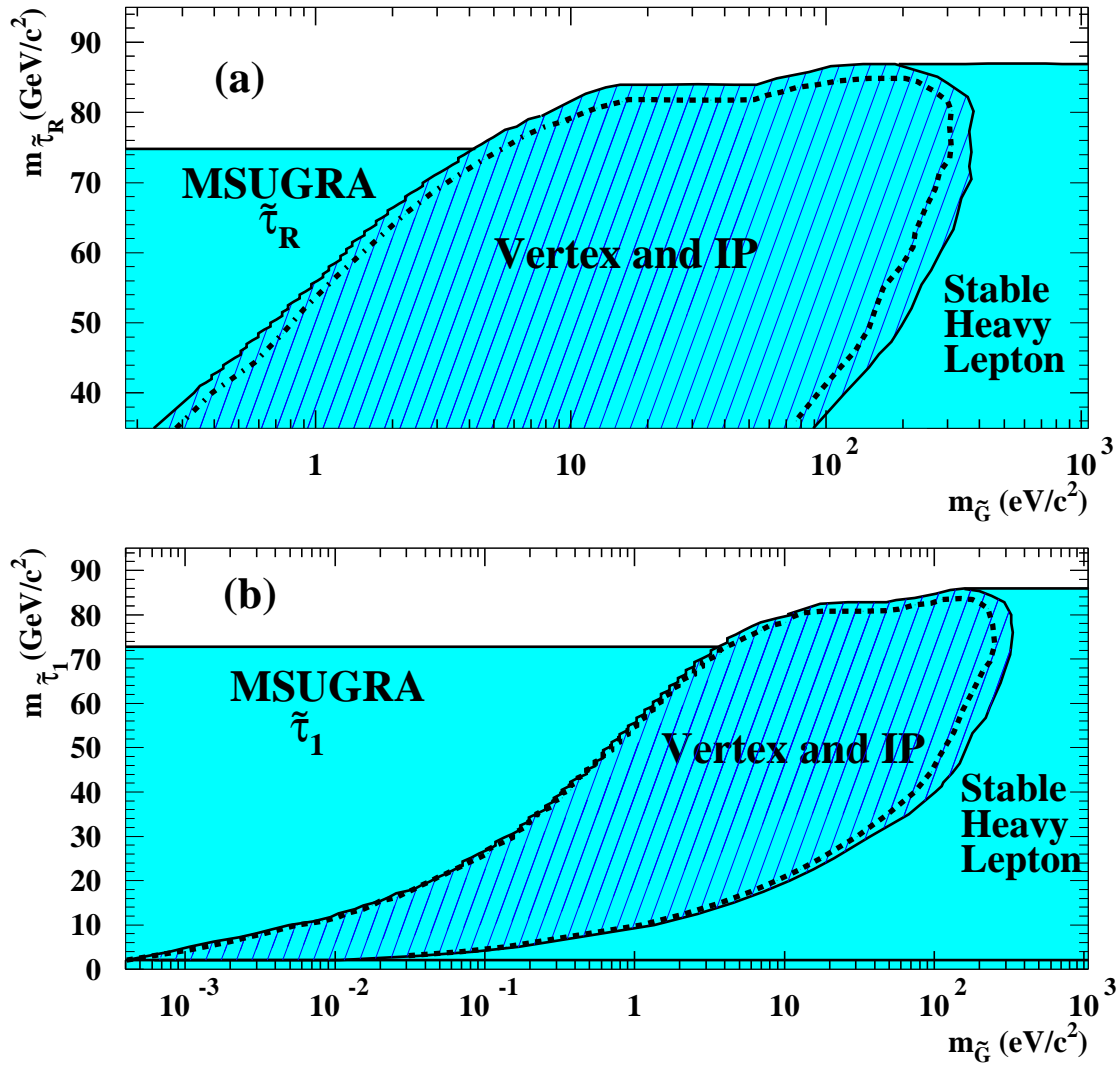}}
  \caption[]{ 
    Exclusion region in the 
    ($m_{\tilde{G}}$,$m_{\tilde{\tau}_R}$) plane (a) and 
    ($m_{\tilde{G}}$,$m_{\tilde{\tau}_1}$) plane (b) 
    at 95\%~C.L. for the present analysis combined 
    with the stable heavy
    lepton search and the search for $\tilde{\tau}_1$ within MSUGRA models, 
    using all LEP-2
    data up to 189 GeV. 
    The positive-slope hatched area shows the region excluded 
    by the impact parameter and secondary vertex searches.
    The dashed line shows the expected limits. The area below the horizontal 
    line in (b) is excluded by the JADE collaboration. A narrow band at 
    $m_{\tau} < m_{\tilde{\tau}} < 2$\GeVcc\ and $m_{\tilde{G}} < 0.03$~\eVcc\
    is not excluded.
} 
  \label{fig:stau1-mass}
\end{figure}

The $\tilde{\tau}_1$ pair production cross-section depends on the mixing in
the stau sector. Therefore, in order to put limits to the $\tilde{\tau}_1$
mass the mixing angle has to be fixed. The results presented here corresponds
to two cases within the $\tilde{\tau}_1$-NLSP scenario. 
In the first case, it is assumed that there is no mixing between the 
\staur~and \staul. Thus, $\tilde{\tau}_1$ is a pure right-handed state
(figure~\ref{fig:stau1-mass}-a). The second case 
(figure~\ref{fig:stau1-mass}-b), corresponds to a mixing angle
which gives the minimum $\tilde{\tau}_1$ pair production cross-section, while
at the same time holds $m_{\tilde{\tau}_1}^2~>$ 0. Since this angle is
close to the region where the coupling to the $Z^0$ almost vanishes, no limit
can be inferred from the LEP1 measurements,
and the search was extended down to stau masses around 2~\GeVcc. 
Stable or long lived particles with masses down to 2~\GeVcc\ are
excluded by the search for heavy stable and long-lived particles in
DELPHI~\cite{heavyparticles}.
Stable or long lived particles with masses below 2~GeV/$c^2$ are 
excluded by the JADE collaboration~\cite{jade}. It is 
assumed that a stau with lower mass than a tau is stable or very 
long lived.
In the case of short lived staus ($m_{\tilde{G}}\lesssim 0.03$~\eVcc), a 
narrow band at $m_{\tau}<m_{\tilde{\tau}_1}<2$~\GeVcc\ is not excluded.
Above 2\GeVcc\ the results from the impact parameter and secondary vertex 
analyses are used for exclusion purposes.

 The impact parameter and secondary 
vertex analyses allow for the exclusion of $\tilde{\tau}_1$
($\tilde{\tau}_R$) 
with a mass below 80\GeVcc\ for gravitino masses between 10 and 310\eVcc\ (8
and 380\eVcc) at 95\%~C.L..
For \mgrav\ 
below a few \eVcc, $m_{\tilde{\tau}_1} < 7$3\GeVcc\ were excluded
by the search for \stuno\ in gravity mediated models~\cite{slep_189}. Results from both searches were not combined because 
the impact parameter searches cover in
excess the overlaping region.
For $m_{\tilde{G}}$ larger than 1000\eVcc\
the limit is 87\GeVcc, obtained after 
combining the results presented in this paper with those of the 
stable heavy lepton search~\cite{heavyparticles}.

\begin{figure}[htbp]
\centerline{\epsfxsize=16.0cm \epsfbox{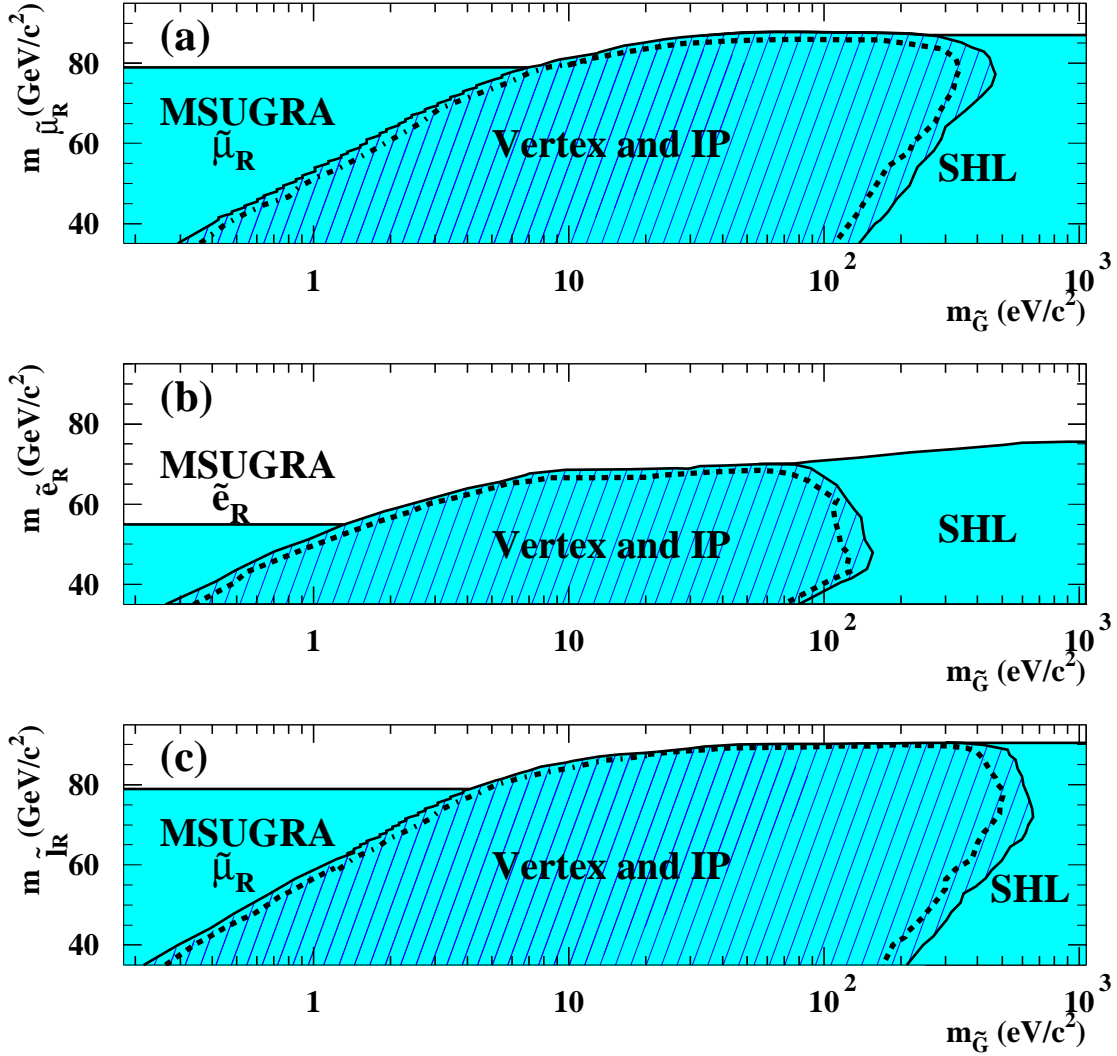}}
  \caption[]{ 
    Exclusion regions in the 
    ($m_{\tilde{G}}$,$m_{\tilde{\mu}_R}$) (a),
    ($m_{\tilde{G}}$,$m_{\tilde{e}_R}$) (b) and
    ($m_{\tilde{G}}$,$m_{\tilde{l}_R}$) (c)
     planes at 95\%~C.L. for the present analyses combined 
    with the stable heavy
    lepton search and the search for $\tilde{\tau}_R$ in gravity mediated
    models, using all LEP2 data up to 189 GeV. 
    The positive-slope hatched area shows the region excluded 
    by the combination
    of the impact parameter and secondary vertex searches.
    The dashed line shows the expected limits.
} 
  \label{fig:slep-mass}
\end{figure}

Within the sleptons co-NLSP scenario, the cross-section limits of 
figures~\ref{fig:cross_todos}-b and -c were used to derive
limits for \smur\ (fig~\ref{fig:slep-mass}-a)
and \selr\ (fig.~\ref{fig:slep-mass}-b) masses at
95\%~C.L.. The $\tilde{\mu}_R$- pair production cross-section is model
independent since it only takes place through the exchange of a $Z^0$ or a
$\gamma$ in the $s$--channel. The $\tilde{e}_R$- pair production
cross-section, however, is a 
function of the GMSB parameters due to the exchange of a
\nuno~in the $t$--channel.
Therefore, in order to put limits to the
\selr\ mass, the aforementioned region of the GMSB parameter space 
was scanned, and for
each selectron mass the smallest theoretical production cross-section was 
chosen for comparison with the experimental limits. 

The t-channel interference causes a bigger fraction of selectrons to be 
produced in the forward region. This results in a loss of efficiency in
the vertex and stable slepton analyses, that was taken into acount for the
calculation of the limits that are shown in fig.~\ref{fig:slep-mass}-b.

Therefore, within the co-NLSP scenario, the impact parameter
search and the secondary vertex search allow for the exclusion of
$\tilde{\mu}_R$ masses  
below 80\GeVcc\ for gravitino masses between 8 and 450\eVcc.
In the case of 
  $\tilde{e}_R$, masses below 67\GeVcc\ for gravitino masses between 10 and
  80\eVcc\ 
are excluded.

Assuming mass degeneracy between the staus and smuons, 
(fig.~\ref{fig:slep-mass}-c), these searches exclude at 95\% C.L. 
$\tilde{l}_R$\
masses below 84\GeVcc\ for $\tilde{G}$ masses between 9 and 570\eVcc.
For very short lifetimes only
$\tilde{\mu}_R$ was considered since it is the best limit that can be
achieved in absence of slepton combination. 
For $\tilde{G}$ larger than 1000\eVcc\
the limit was 87\GeVcc, obtained from the
stable heavy lepton search~\cite{heavyparticles}. $\tilde{l}_R$\
masses below 35\GeVcc\ are excluded from LEP~1 data~\cite{lep1ex}.
In the case of $\tilde{l}_R$\ degeneracy, this limit improves to 41\GeVcc.


\subsubsection{Chargino pair production}

The limits on chargino pair production cross-section 
were used to exclude areas within the 
($m_{\tilde{\chi}_1^+},m_{\tilde{\tau}_1}$) plane in different 
domains of the gravitino mass~\cite{Dutta}.

\begin{figure}[htbp]
\epsfxsize=16.0cm
\centerline{\epsffile{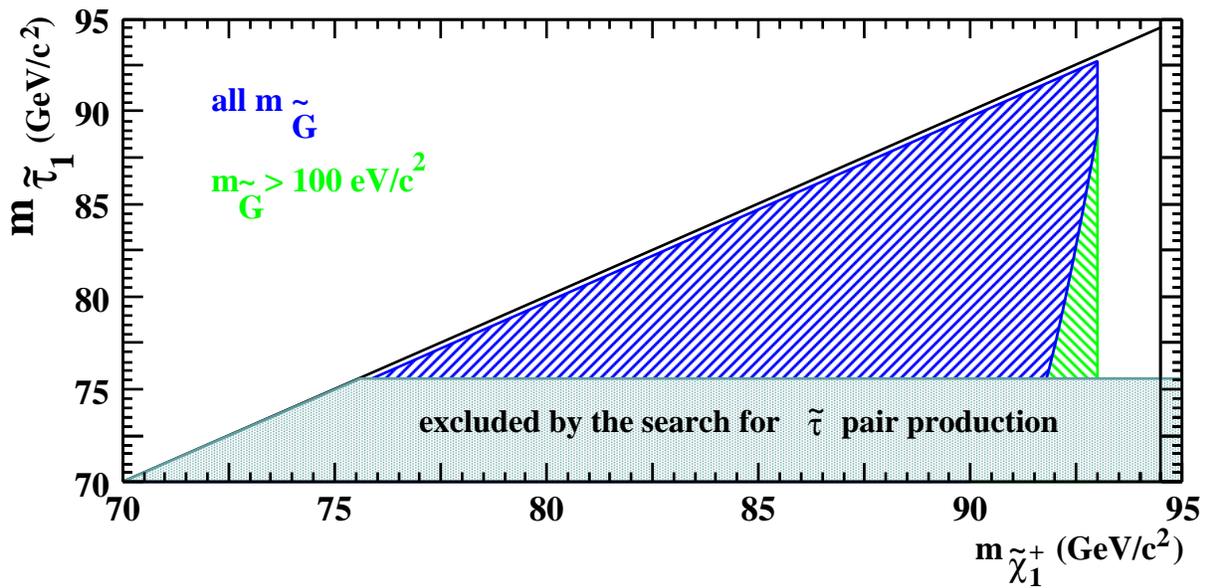}}
\caption[]{Areas excluded at 95\% CL in the 
($m_{\tilde{\chi}_1^+}$,$m_{\tilde{\tau}_1}$) plane.
The positive-slope
area is excluded for all $m_{\tilde{G}}$.
The negative-slope area is excluded only for  
$m_{\tilde{G}} \geq 1$00\eVcc . The grey area is excluded 
by the direct search for stau pair production~\cite{slep_189}.}
\label{fig:mass}
\end{figure}

Figure~\ref{fig:mass} shows the regions excluded at 
95\% CL in the 
($m_{\tilde{\chi}_1^+}$,$m_{\tilde{\tau}_1}$) plane.
The positive-slope area is excluded for all gravitino masses. The 
negative-slope area is only excluded for $m_{\tilde{G}} \geq 100$\eVcc.
The area below $m_{\tilde{\tau}_1} = 75.8$\GeVcc\ is excluded
by the direct search for stau pair production~\cite{slep_189}.
The area of $\Delta m  \leq  0.3$\GeVcc\ is not excluded because 
in this regions the charginos do not decay mainly 
in \stuno\ and $\nu_{\tau}$, but in W and $\tilde{G}$.
Thus, if 
$\Delta m\! \geq\! 0.3$\GeVcc , 
limits at 91.8, 93.0 and $93.0$\GeVcc\ 
can be set for $m_{\tilde{G}} = 1$, 100 and 
1000\eVcc\ respectively.
The limit at 
$m_{\tilde{G}}\! =\! 1$\eVcc\ is also valid for 
smaller masses of the gravitino, because they lead to the same final state 
topologies. The same argument is true for 
$m_{\tilde{G}} \geq\! 1$\keVcc.
The chargino mass limit 
decreases with decreasing $m_{\tilde{\tau}_1}$\   because 
in scenarios with gravitino LSP, small stau masses correspond to 
small sneutrino masses (both are proportional to $\Lambda$), 
and hence to smaller 
production 
cross-sections 
due to the destructive interference between the 
$s$- and $t$-channels.
It should be noticed that within the parameter space that concerns this work, 
the lightest chargino is at least 40\% heavier than the lightest neutralino. 
Thus, for small gravitino masses the search for neutralinos implies a lower 
limit
on the lightest chargino of 120\GeVcc. Neutralinos are not directly searched 
for in heavier gravitino mass regions and therefore the limit of 93\GeVcc\ 
remains valid.

\subsubsection{Limits on the GMSB parameter space} 

\begin{figure}[htbp]
\epsfxsize=16.0cm
\centerline{\epsffile{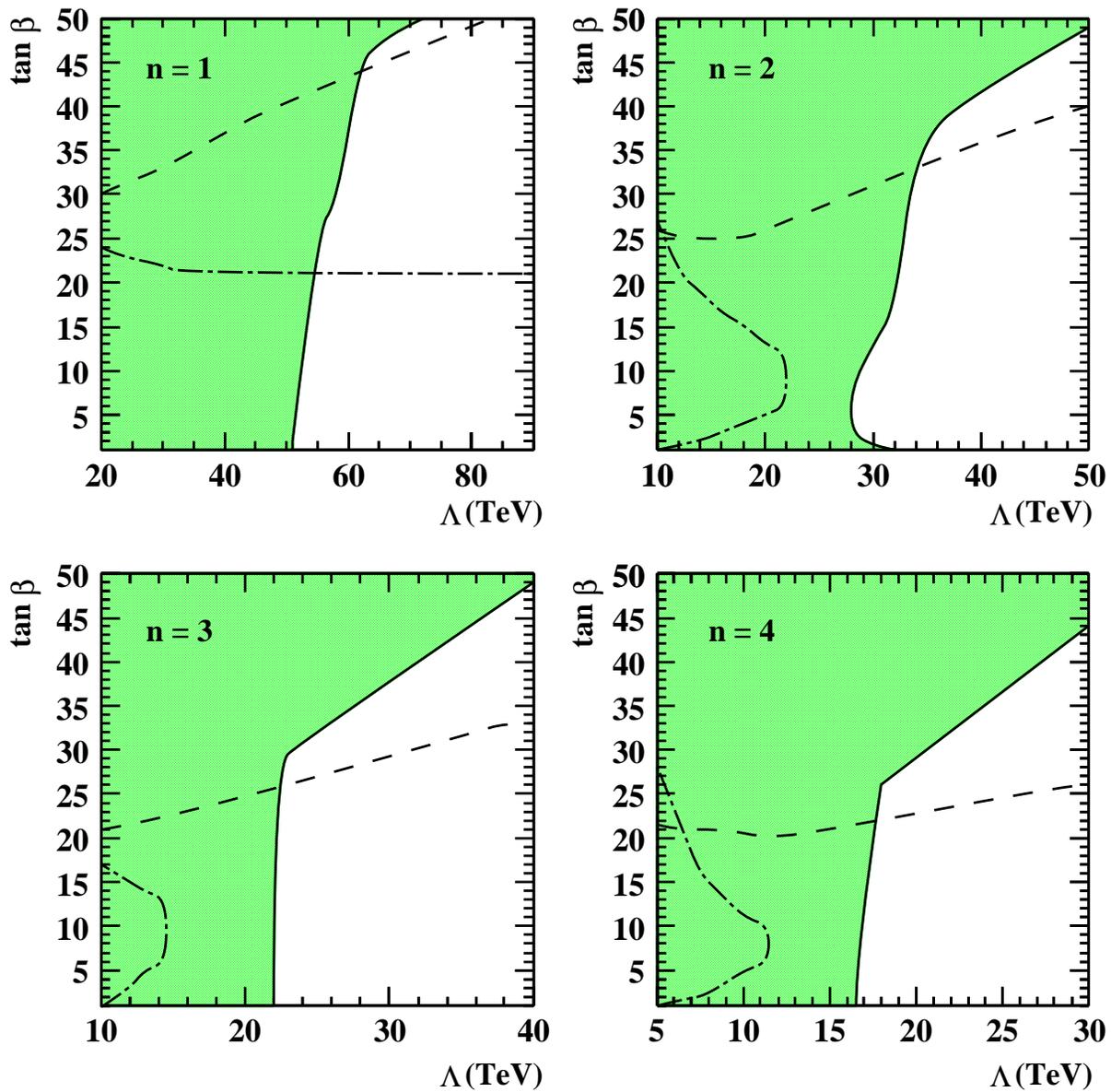}}
\caption{Shaded areas in the 
($\tan\beta , \Lambda$) plane are excluded at 95\% CL.
The areas below 
the dashed lines contain points of the GMSB parameter space with \nuno -NLSP.
The areas to the right (above for $n = 1$) of the dashed-dotted lines 
contain points of the GMSB parameter space were sleptons are the NLSP.}
\label{fig:lambda}
\end{figure}

Finally, all these results can be combined to produce exclusion plots 
within the ($\tan\beta, \Lambda$) space. 
As an example,  fig.~\ref{fig:lambda} shows the zones
 excluded for $n=$1 to 4 for $m_{\tilde{G}} \leq 1$\eVcc , which corresponds to
the NLSP decaying at the main vertex. 
The shaded areas are excluded. The areas below 
the dashed lines contain points of the GMSB parameter space with \nuno -NLSP.
The areas to the right (above for $n = 1$) of the dashed-dotted lines 
contain points of the GMSB parameter space where sleptons are the NLSP. 
It can be seen that the region of slepton-NLSP increases with $n$. 
The contrary occurs to the region of neutralino-NLSP. 
A limit can be set for the variable $\Lambda$\ at 16.5~TeV.

\section{Summary}
Lightest neutralino-, slepton- and chargino pair production 
were searched for in the 
context of light gravitino scenarios. Two scenarios
were explored: the $\tilde{\tau}_1$ NLSP  and the $\tilde{l}_R$ co-NLSP
scenarios.  No evidence for signal production was found.
Hence, the DELPHI collaboration 
sets lower limits at 95\% C.L. for the mass of 
the $\tilde{\chi}_1^0$ at 82\GeVcc\ if $m_{\tilde{G}}\leq 1$\eVcc, 
for the mass of 
the $\tilde{\tau}_1$ at 73\GeVcc, the $\tilde{l}_R$ at 79\GeVcc , 
and the lightest chargino at 93\GeVcc\ for all $m_{\tilde{G}}$.
A limit is also set on the variable $\Lambda$\ at 16.5 TeV 
if $m_{\tilde{G}}\leq 1$\eVcc .

\subsection*{Acknowledgements}
\vskip 3 mm
 We are greatly indebted to our technical 
collaborators, to the members of the CERN-SL Division for the excellent 
performance of the LEP collider, and to the funding agencies for their
support in building and operating the DELPHI detector.\\
We acknowledge in particular the support of \\
Austrian Federal Ministry of Science and Traffics, GZ 616.364/2-III/2a/98, \\
FNRS--FWO, Belgium,  \\
FINEP, CNPq, CAPES, FUJB and FAPERJ, Brazil, \\
Czech Ministry of Industry and Trade, GA CR 202/96/0450 and GA AVCR A1010521,\\
Danish Natural Research Council, \\
Commission of the European Communities (DG XII), \\
Direction des Sciences de la Mati$\grave{\mbox{\rm e}}$re, CEA, France, \\
Bundesministerium f$\ddot{\mbox{\rm u}}$r Bildung, Wissenschaft, Forschung 
und Technologie, Germany,\\
General Secretariat for Research and Technology, Greece, \\
National Science Foundation (NWO) and Foundation for Research on Matter (FOM),
The Netherlands, \\
Norwegian Research Council,  \\
State Committee for Scientific Research, Poland, 2P03B06015, 2P03B1116 and
SPUB/P03/178/98, \\
JNICT--Junta Nacional de Investiga\c{c}\~{a}o Cient\'{\i}fica 
e Tecnol$\acute{\mbox{\rm o}}$gica, Portugal, \\
Vedecka grantova agentura MS SR, Slovakia, Nr. 95/5195/134, \\
Ministry of Science and Technology of the Republic of Slovenia, \\
CICYT, Spain, AEN96--1661 and AEN96-1681,  \\
The Swedish Natural Science Research Council,      \\
Particle Physics and Astronomy Research Council, UK, \\
Department of Energy, USA, DE--FG02--94ER40817. \\

\newpage
%

\begin{table}[hbtp]
  \begin{center}
    \begin{tabular}{||c||c|c|c|c||c||c||}
      \hline
 Cut & $\gamma\gamma$ &$f\bar{f}\gamma$ & 4-fermion& Total MC & Data
 &                       Signal \\
      \hline
pre-selection &$1134\pm 37 $&$413\pm 14$&$385\pm 18$&$1933\pm43$&1791&59.5\%\\
1          &$28\pm 3   $&$ 330\pm 11 $ &$363\pm 18$ &$721.3\pm21$&706 &51.2\%\\
2          &$4.3\pm 1.2$&$11.1\pm 1.5 $&$12.5\pm2.8$ &$28.0\pm3.4$&24 &42.1\%\\
3          & 0          &$0.07\pm 0.07$&$1.13\pm0.3$&$1.2\pm0.3$&2 &32.8\%\\
      \hline
    \end{tabular}
  \end{center}
  \caption[]{Number of events remaining in the data and simulated samples
at $\sqrt{s} = 189$ \GeV\ after the various stages of the selection procedure
described in the search for neutralinos decaying into slepton and lepton. 
The signal efficiencies corresponds to  
$m_{\tilde{\chi}^0_1}=87~$\GeVcc\ and 
$m_{\tilde{l}}=75~$\GeVcc.}
  \label{tab:evol}
  \end{table}
\begin{table}[hbtp]
  \begin{center}
    \begin{tabular}{||c||c|c|c|c||c||c||}
      \hline
 Cut & $\gamma\gamma$ &$f\bar{f}\gamma$ & 4-fermion& Total MC & Data
 &                       Signal \\
      \hline
pre-selection &$1134\pm 37 $&$413\pm 14$&$385\pm 18$&$1933\pm43$&1791& 57.7\%\\
1&$66.5\pm 7$&$ 376.4\pm 13.0 $ &$331.0\pm 12.1$ &$468.5\pm 19.1$&404 &54.5\%\\
2          &$6.7\pm 1.8$&$9.5\pm 1.3 $&$10.6\pm0.9$ &$26.8\pm2.4$&23 &44.9\%\\
3          & 0          &$0.07\pm 0.07$&$1.09\pm0.3$&$1.16\pm0.3$&1 &37.3\%\\
      \hline
    \end{tabular}
  \end{center}
  \caption[]{Number of events remaining in the data and simulated samples
at $\sqrt{s} = 189$ \GeV\ after the various stages of the selection procedure
described in the search for neutralinos decaying into stau and tau. 
The signal efficiencies corresponds to  
$m_{\tilde{\chi}^0_1}=87~$\GeVcc\ and 
$m_{\tilde{\tau}_1}=75~$\GeVcc.}
  \label{tab:evol-tau}
  \end{table}

\begin{table}[hbt]
\begin{center}
\begin{tabular}{||c|c||} \hline 
Observed events
             & 1                                            \\ \hline
Total background
     & 1.42$^{+0.72}_{-0.36}$                              \\ \hline    
\hline
$Z^*/\gamma \to (ll) (n\gamma)$
     & 0.23$^{+0.35}_{-0.01}$ \\ \hline
4-fermion (except $\gamma\gamma$)
   & 0.45$\pm 0.16$\\ \hline
$\gamma\gamma \to \tau^+\tau^-$
    & 0.74$^{+0.59}_{-0.32}$\\ \hline
\end{tabular}
\end{center}
\caption[.]{
Number of observed events at $\sqrt{s}=189$~GeV,
together with the total number of expected background events
and the expected numbers from the individual background sources,
for both large impact parameter and secondary vertex searches combined.}
\label{tab:grav:bkg}
\end{table}

\begin{table}[hbtp]
  \begin{center}
    \begin{tabular}{||c|c||}
      \hline
 Observed events & $ 4 $ \\
\hline
 Total background & $ 4.54^{+1.12}_{-0.57}$ \\ \hline \hline
 $Z^*/\gamma \to (\tau \tau) (n\gamma)$  & $1.33^{+0.46}_{-0.35}$  \\
\hline
 $\gamma \gamma \rightarrow \tau^+ \tau^-$ & $0.61^{+0.99}_{-0.38}$   \\
\hline
 WW                                    &
  $2.52^{+0.26}_{-0.23}$ \\
\hline
 ZZ                                    &
 $ 0.08^{+0.04}_{-0.03}$\\
\hline
    \end{tabular}
  \end{center}
  \caption{Expected simulated SM background events and selected data 
events at 189 \GeV\ centre-of-mass energy for the small 
impact parameter search.}
  \label{tab:cavallo1}
  \end{table}

\begin{table}[hbtp]
  \begin{center}
    \begin{tabular}{|c||c|c|c|c|}
      \hline
Sample & Efficiencies (\%) & Main backgrounds & Expected b.g. 
& Observed events \\ \hline \hline
$m_{\tilde{G}}=1$\eVcc & 24 - 36 &WW , $\gamma\gamma$ &38.9$\pm 4.9$ & 36 
\\ \hline
$m_{\tilde{G}}=100$\eVcc & 28 - 50 & $\gamma\gamma$ &2.1$\pm 0.9$ & 1 
\\ \hline
$m_{\tilde{G}}=1000$\eVcc & 0 - 63&$\mu\mu(\gamma)$ & 1.7$\pm 0.3$ & 1
\\ \hline
    \end{tabular}
  \end{center}
  \caption[] {Range of efficiencies for the different sets of chargino signals 
described in section~\ref{experimentalprocedure}, main sources of background, 
expected background and observed data events for the different analyses.}
  \label{tab:results}
  \end{table}

\begin{table}[hbtp]
  \begin{center}
    \begin{tabular}{||c|c|c||}
      \hline
 $n$ &co-NLSP(\GeVcc ~) & \stuno -NLSP (\GeVcc ~)  \\
      \hline
1 & 83.0 & 82.5 \\
2 & 85.0 & 85.0 \\
3 & 86.0 & 86.0 \\
4 & 87.0 & 87.0 \\
      \hline
    \end{tabular}
  \end{center}
  \caption{The 95\% C.L. lower limits on $m_{\tilde{\chi}_1^0}$
within the \stuno -NLSP and co-NLSP scenarios for different $n$.}
  \label{tab:mass}
  \end{table}

\clearpage



\vspace{1cm}

\clearpage

\end{document}